\documentclass[11pt,oneside,onecolumn,draftclsnofoot]{IEEEtran}
\usepackage[dvips]{graphicx}
\usepackage{epsfig}
\usepackage{color}
\usepackage{amsmath,amssymb,amsfonts,cite}
\usepackage{amsthm}
\def\xv{{\mathbf x}}
\def\yv{{\mathbf y}}
\def\zv{{\mathbf z}}
\def\hv{{\mathbf h}}
\def\gv{{\mathbf g}}
\def\Xv{{\mathbf X}}
\def\Yv{{\mathbf Y}}
\def\Zv{{\mathbf Z}}
\def\pv{{\mathbf p}}
\def\Pc{{\mathcal{P}}}
\def\Lc{{\mathcal L}}
\def \Pr {{\mathrm{Pr}}}
\def\Cc{{\mathcal{C}}}

\def\Wc{{\mathcal{W}}}
\def\Mc{{\mathcal{M}}}
\def\E{{\mathbb{E}}}

\theoremstyle{plain}
\newtheorem{theorem}{Theorem}

\newtheorem{lemma}{Lemma}
\newtheorem{corollary}{Corollary}
\newtheorem*{lemmaC1}{Lemma C.1}
\newtheorem*{lemmaD1}{Lemma D.1}
\newtheorem*{lemmaD2}{Lemma D.2}

\newtheorem*{lemmaA1}{Lemma A.1}

\title{\huge {On the Throughput of Secure Hybrid-ARQ Protocols \\for Gaussian Block-Fading Channels}}

\author{Xiaojun~Tang,~Ruoheng~Liu,~Predrag~Spasojevi\'{c},~and~H.~Vincent~Poor
\thanks{This research was supported by the National Science Foundation
under Grants ANI-03-38807, CNS-06-25637 and CCF-07-28208. The material in this
paper was presented in part at the IEEE International Symposium on
Information Theory, Nice, France, June 24 - 29, 2007.}%
\thanks{X. Tang and P. Spasojevi\'{c} are with Wireless Information Network Laboratory (WINLAB), Department of Electrical and Computer Engineering, Rutgers University,
North Brunswick, NJ 08902, USA (e-mail:
\{xtang,spasojev\}@winlab.rutgers.edu).}%
\thanks{R. Liu and H. V. Poor are with Department of Electrical Engineering, Princeton
University, Princeton, NJ 08544, USA (email:
\{rliu,poor\}@princeton.edu).}}

\begin{document}

\maketitle

\IEEEpeerreviewmaketitle

\begin{abstract}
The focus of this paper is an information-theoretic study of
retransmission protocols for reliable packet communication under a
secrecy constraint. The \emph{h}ybrid \emph{a}utomatic
\emph{r}etransmission re\emph{q}uest (HARQ) protocol is revisited
for a block-fading wire-tap channel, in which two legitimate users
communicate over a block-fading channel in the presence of a passive
eavesdropper who intercepts the transmissions through an independent
block-fading channel. In this model, the transmitter obtains a
$1$-bit ACK/NACK feedback from the legitimate receiver via an
error-free \emph{public} channel. Both reliability and
confidentiality of secure HARQ protocols are studied by the joint
consideration of channel coding, secrecy coding, and retransmission
protocols. In particular, the error and secrecy performance of
\emph{repetition time diversity} (RTD) and \emph{incremental
redundancy} (INR) protocols are investigated based on \emph{good}
Wyner code sequences, which ensure that the confidential message is
decoded successfully by the legitimate receiver and is kept in total
ignorance by the eavesdropper for a given set of channel
realizations. This paper first illustrates that there exists a
\emph{good} rate-compatible Wyner code family which ensures a secure
INR protocol. Next, two types of outage probabilities,
\emph{connection outage} and \emph{secrecy outage} probabilities are
defined in order to characterize the tradeoff between the
reliability of the legitimate communication link and the
confidentiality with respect to the eavesdropper's link. For a given
connection/secrecy outage probability pair, an achievable throughput
of secure HARQ protocols is derived for block-fading channels.
Finally, both asymptotic analysis and numerical computations
demonstrate the benefits of HARQ protocols to throughput and
secrecy.

\newpage

\end{abstract}

\section{Introduction}\label{sec:intro}

Reliable communication is essential in applications of wireless
packet-oriented data networks. A class of special coding schemes,
the so-called hybrid automatic retransmission request (HARQ),
combine powerful channel coding with retransmission protocols to
enhance the reliability of communication links. Among currently
available HARQ protocols, the most elementary form is the {\it
repetition-coding-based} HARQ which combines several noisy
observations of the same packet by using a suitable diversity
technique at the receiver, such as maximal-ratio combining,
equal-gain combining, or selection combining. A more powerful HARQ
scheme is the so-called {\it incremental redundancy} HARQ, which
achieves higher throughput efficiency by adapting its error
correcting code redundancy to fluctuating channel conditions. In an
incremental redundancy scheme, the message is encoded at the
transmitter by a ``mother'' code. Initially, only a selected number
of coded symbols are transmitted. The selected number of coded
symbols form a codeword of a punctured mother code. If a
retransmission is requested, additional redundancy symbols are sent
under possibly different channel conditions. An
information-theoretic analysis of the throughput performance of HARQ
protocols over block-fading Gaussian collision channels is found in
\cite{Caire:IT:01}. By assuming Gaussian random coding and
typical-set decoding, the results of \cite{Caire:IT:01} are
independent of the particular coding/decoding technique and can be
regarded as providing a limiting performance in the
information-theoretic sense. Another line of recent research on HARQ
concerned with various mother codes and their puncturing can be
found in
\cite{Hagenauer:TC:88,Narayanan:CL:97,Tuninetti:IT:02,emina:DIMACS,Sesia:TC:04,
Leanderson:TWC:04,Soljanin:IT:05}.

Confidentiality is a basic requirement for secure communication over
wireless networks. We note that the broadcast nature of the wireless
medium gives rise to a number of security issues. In particular,
wireless transmission is very susceptible to eavesdropping since
anyone within communication range can listen to the traffic and
possibly extract information. Traditionally, confidentiality has
been provided by using cryptographic methods, which rely heavily on
secret keys. However, the distribution and maintenance of secret
keys are still open issues for large wireless networks. Fortunately,
confidential communication is possible without sharing a secret key
between legitimate users. This was shown by Wyner in his seminal
paper \cite{Wyner:BSTJ:75}. In the discrete memoryless wire-tap
channel model he proposed, the communication between two legitimate
users is eavesdropped upon via a degraded channel (the eavesdropper
channel). The level of ignorance of the eavesdropper with respect to
the confidential message is measured by the equivocation rate.
Perfect secrecy requires that the equivocation rate should be
asymptotically equal to the message entropy rate. Wyner showed that
perfect secrecy can be achieved via a stochastic code, referred to
as Wyner secrecy code. Csisz{\'{a}}r and K{\"{o}}rner generalized
this result and determined the secrecy capacity region of the
broadcast channel with confidential messages in
\cite{Csiszar:IT:78}. Recent research investigates multi-user
communication with confidential messages, e.g., multiple access
channels with confidential messages \cite{Liang:IT:06,Liu:ISIT:06},
multiple access wire-tap channels \cite{Tekin:ISIT:06}, and
interference channels with confidential messages \cite{Liu07it}. The
effect of fading on secure communication has been studied in
\cite{Barros:ISIT:06,Liang06novit,Li:Allerton:06,Gopala:IT:06}. More
specifically, assuming that all communicating parties have perfect
channel state information (CSI) prior to the message transmission,
\cite{Barros:ISIT:06} has studied the delay limited secrecy capacity
of wireless channels, while
\cite{Liang06novit,Li:Allerton:06,Gopala:IT:06} have studied the
secrecy capacity of an ergodic fading channel. \cite{Gopala:IT:06}
has also considered the ergodic scenario in which the transmitter
has no CSI about the eavesdropper channel.

In this paper, we investigate secure packet communication based on
HARQ protocols. The challenge of this problem is twofold: first, the
encoder at the transmitter needs to provide sufficient redundancy
for the legitimate receiver to decode its message successfully; on
the other hand, too much redundancy may help adversarial
eavesdropping. As an example, retransmission is an effective way to
enhance reliability, but nevertheless it may also compromise
confidentiality. This motivates the joint consideration of channel
coding, secrecy coding, and retransmission protocols.

We consider a frequency-flat block-fading Gaussian wire-tap channel.
In this model, a transmitter sends confidential messages to a
legitimate receiver via a block-fading channel in the presence of a
passive eavesdropper who intercepts the transmission through an
independent block-fading channel. We assume that the transmitter has
no perfect CSI, but receives a $1$-bit ACK/NACK feedback from the
legitimate receiver via a reliable public channel. Under this
setting, we study the secure HARQ protocols from an information
theoretic point of view. In particular, the error and secrecy
performance of \emph{repetition time diversity} (RTD) and
\emph{incremental redundancy} (INR) protocols are investigated based
on \emph{good} Wyner code sequences, which ensure that the
confidential message is decoded successfully by the legitimate
receiver and is kept completely secret from the eavesdropper for a
given set of channel realizations of both the main and the
eavesdropper channels. Next, we show that there exists a \emph{good}
rate-compatible Wyner code family which suits the secure INR
protocol. Due to the absence of CSI, the transmitter cannot adapt
its code and power level to channel conditions. Instead, for a given
mother code, we consider the outage performance of secure HARQ
protocols. Specifically, we define two types of outage:
\emph{connection outage} and \emph{secrecy outage}. The outage
probabilities (i.e., the probabilities of connection and secrecy
outage) are used to characterize the tradeoff between the
reliability of the legitimate communication link and the
confidentiality with respect to the eavesdropper's link. We evaluate
the achievable throughput of HARQ protocols under the constraints on
these two outage probabilities. Finally, we compare the secrecy
throughput of two HARQ protocols through both numerical computations
and an asymptotic analysis, and illustrate the benefit of HARQ
schemes to information secrecy.

Generally speaking, when the coding parameters (main channel code
rate and secrecy information rate for ensuring reliability and
secrecy, respectively) can be freely chosen, INR can achieve a
significantly larger throughput than RTD, which concurs with the
results not involving secrecy where it has been shown that
mutual-information accumulation (INR) is a more effective approach
than SNR-accumulation (RTD) \cite{Caire:IT:01}. However, when one is
forced to ensure small connection outage for the main channel even
when it is bad, one is forced to reduce the main channel code rate.
The INR scheme, having a larger coding gain (to both the intended
receiver and the eavesdropper), needs to sacrifice a larger portion
of the main channel code rate in order to satisfy the secrecy
requirement. Hence, when the main channel code rate is bounded due
to the connection outage constraint, the achievable secrecy
throughput of INR may be smaller than that of RTD. This result
deviates from that not involving secrecy.

The remainder of this paper is organized as follows. We describe the
system model and preliminaries in Section \ref{sec:model}. In
Section~\ref{sec:coding}, we prove the existence of good Wyner codes
for parallel channel communication and define outage events, while
these results are applied to INR and RTD protocols in Section
\ref{sec:sHARQ}. We derive the secrecy throughput of two protocols
over block fading channels in Section \ref{sec:secthru}, and present
an asymptotic analysis in Section \ref{sec:limit}. We illustrate and
compare the various results and protocols numerically in
Section~\ref{sec:sac}. Finally, we give conclusions and some
interesting directions for future research in
Section~\ref{sec:conclusions}, The proofs of the results are
provided in appendices.


\section{System Model and Preliminaries}\label{sec:model}

\begin{figure}[bt]
  \includegraphics[width=0.8\linewidth]{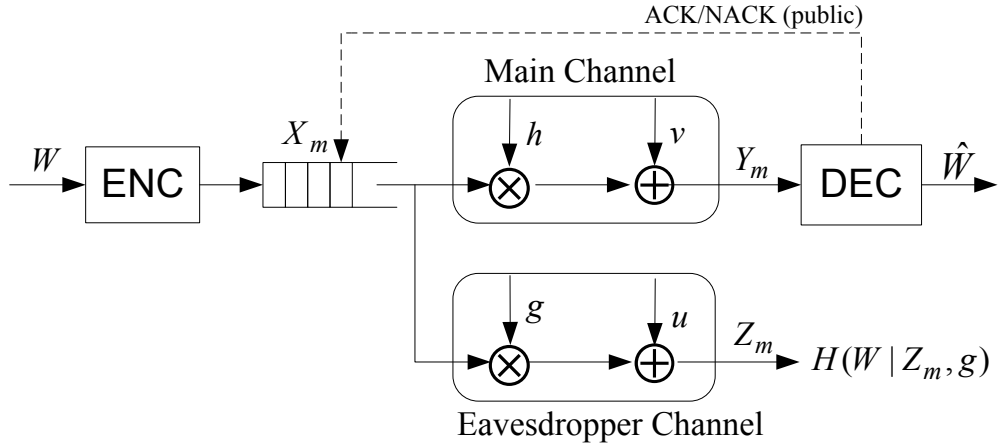}\\
  \centering
  \caption{System model: hybrid ARQ protocols for the block-fading channel in the presence of a passive eavesdropper}\label{channel}
\end{figure}

\subsection{System Model}\label{sec:model:channel}

As shown in Fig.~\ref{channel}, we consider a model in which a transmitter
sends confidential messages to a destination via a source-destination channel
(the main channel) in the presence of a passive eavesdropper which listens to
the transmission through a source-eavesdropper channel (the eavesdropper
channel). Both the main channel and the eavesdropper channel experience
$M$-block fading, in which the channel gain is constant within a block while
varying independently from block to block \cite{Shamai:VT:94,bigl:IT:98}. We
assume that each block is associated with a time slot of duration $T$ and
bandwidth $W$; that is, the transmitter can send $N=\lfloor 2WT\rfloor$ real
symbols in each slot. Additionally, we assume that the number of channel uses
within each slot (i.e., $N$) is large enough to allow for invoking random
coding arguments.\footnote{For example, in a 64 kb/s down-link reference data
channel for universal mobile telecommunications system (UMTS) data-transmission
modes, each slot can contain up to $N \approx 10000$ dimensions
\cite{Holma:02}.}

At the transmitter, a confidential message $w \in \Wc$ is encoded
into a codeword $x^{MN}$, which is then divided into $M$ blocks
$[x_1^N,x_2^N,\dots,x_M^N]$, each of length $N$. The codeword
$x^{MN}$ occupies $M$ slots; that is, for $i=1,\dots,M$, the $i$-th
block $x_i^N$ is sent in slot $i$ and received by the legitimate
receiver through the channel gain $h_i$ and by the eavesdropper
through the channel gain $g_i$. A discrete time baseband-equivalent
block-fading wire-tap channel model can be expressed as follows:
\begin{align}
&           &  y(t) &= \sqrt{h_i}x(t) + v(t) \nonumber&\\
&\text{and} &  z(t) &= \sqrt{g_i}x(t) + u(t)
\quad\text{for}~t=1,\dots,MN,~i=\left\lceil
  t/N\right\rceil,   &\label{eq:chm}
\end{align}
where $x(t)$ denotes the input signal, $y(t)$ and $z(t)$ denote the output
signals at the legitimate receiver and the eavesdropper, respectively, at time
$t$ ($t=1,\dots, MN$), $\{v(t)\}$ and $\{u(t)\}$ are independent and
identically distributed (i.i.d.) $\mathcal{N}(0, 1)$ random variable sequences,
and $h_i$ and $g_i$, for $i=1,\dots,M$, denote the normalized (real) channel
gains of the main channel and the eavesdropper channel, respectively.
Furthermore, we assume that the signal ${x(t)}$ has constant average energy per
symbol
\begin{equation}\label{power}
    E[|x(t)|^2] \leq \bar{P}.
\end{equation}

Let $\hv=\left[h_1,\dots, h_M\right]$ and $\gv=\left[g_1,\dots, g_M\right]$
denote vectors whose elements are the main channel gains and the eavesdropper
channel gains, respectively. We refer to $(\hv, \gv)$ as a {\it channel pair}
and assume that the legitimate receiver knows its channel $\hv$, while the
eavesdropper knows its channel $\gv$.

\subsection{Wyner Codes}\label{sec:model:wyner}

In this subsection, we consider a single-block transmission, i.e., $M=1$ and
introduce Wyner codes \cite{Wyner:BSTJ:75}, which are the basis of our secure
HARQ protocols.

Let $C(R_0,R_s,N)$ denote a Wyner code of size $2^{NR_0}$ to convey
a confidential message set $\Wc = \{1, 2, \dots, 2^{N{R_s}}\}$,
where $R_0 \geq R_s$ and  $N$ is the codeword length. The basic idea
of Wyner codes is to use a stochastic encoder to increase the
secrecy level \cite{Wyner:BSTJ:75,Csiszar:IT:78}. Hence, there are
two rate parameters associated with the Wyner code: the main channel
code rate $R_0$ and the secrecy information rate $R_s$.\footnote{We
call $R_0-R_s$ the secrecy gap as the rate sacrificed to ensure the
secrecy requirement.} The Wyner code $C(R_0,R_s,N)$ is constructed
based on random binning \cite{Wyner:BSTJ:75} as follows. We generate
$2^{NR_0}$ codewords $x^{N}(w,v)$, where $w=1,2,\dots, 2^{NR_s}$,
and $v=1,2,\dots,2^{N(R_0-R_s)}$, by choosing the $N2^{NR_0}$
symbols $x_{i}(w,v)$ independently at random according to the input
distribution $p(x)$. A Wyner code ensemble $\Cc(R_0,R_s,N)$ is the
{\it set} of all possible Wyner codes of length $N$, each
corresponding to a specific generation and a specific labeling.

The stochastic encoder of $C(R_0,R_s,N)$ is described by a matrix of
conditional probabilities so that, given $w \in \Wc$, we randomly and uniformly
select $v$ from $\{1,2,\dots,2^{N(R_0-R_s)}\}$ and transmit $x^N=x^{N}(w,v)$.
We assume that the legitimate receiver employs a typical-set decoder. Given
$y^N$, the legitimate receiver tries to find a pair $(\tilde{w},\tilde{v})$ so
that $x^{N}(\tilde{w},\tilde{v})$ and $y^N$ are jointly typical
\cite{Cover:91}, i.e.,
$$\{x^{N}(\tilde{w},\tilde{v}),y^N\} \in T_{\epsilon}^{N}(P_{XY}).$$
If there is no such jointly typical pair, then the decoder claims
failure.

Assume that signals $y^N$ and $z^N$ are received at the legitimate
receiver and the eavesdropper, respectively, via a channel pair
$(h,g)$. The average error probability is defined as
\begin{equation}\label{pe1}
    P_e(h) = \sum_{w \in \Wc} \Pr\left\{\phi\bigl(Y^N(w)\bigr) \neq w | h, w
    \mathrm{~sent} \right\}\Pr(w),
\end{equation}
where $\phi\bigl(Y^N(w)\bigr)$ is the output of the decoder at the legitimate
receiver and $\Pr(w)$ is the prior probability that message $w \in \Wc$ is
sent.

The secrecy level, i.e., the degree to which the eavesdropper is confused, is
measured by the equivocation rate at the eavesdropper. \emph{Perfect secrecy}
is achieved if for all $\epsilon
>0$ the equivocation rate satisfies
\begin{equation}\label{fequivoc1}
    \frac{1}{N}H(W|g,Z^N) \geq \frac{1}{N}H(W)-\epsilon.
\end{equation}

For conciseness, we say that a code $C$ of length $N$ is \emph{good} for a
wire-tap channel with the channel pair $(h,g)$ if $P_e(h) \leq \epsilon$ and
the perfect secrecy requirement (\ref{fequivoc1}) can be achieved, for all
$\epsilon>0$ and sufficiently large $N$.

\subsection{Secure HARQ Protocols}\label{sec:model:sharq}

We first consider a general (in $M$) secure HARQ protocol for a block-fading
wire-tap channel. The transmitter encodes the confidential information (and
cyclic redundancy check (CRC) bits) by using a mother code of length $MN$. The
obtained codeword $x^{MN}$ is partitioned into $M$ blocks represented as
$[x_1^N,x_2^N,\dots,x_M^N]$. At the first transmission, the transmitter sends
the block $x_1^N$ under the channel gain pair $(h_1,g_1)$. Decoding of this
code is performed at the intended receiver, while the secrecy level is measured
at the eavesdropper. If no error is detected, the receiver sends back an
acknowledgement (ACK) to stop the transmission; otherwise a negative
acknowledgement (NACK) is sent to request retransmission, and the transmitter
sends the block $x_2^N$ under the channel gain pair $(h_2,g_2)$. Now, decoding
and equivocation calculation are attempted at the receiver and eavesdropper by
combining the previous block $x_1^N$ with the new block $x_2^N$. The procedure
is repeated after each subsequent retransmission until all $M$ blocks of the
mother code are transmitted or an HARQ session completes due to the successful
decoding at the intended receiver.

Now, we focus on the error performance and secrecy level after $m$
transmissions, $m=1,2,\dots,M$. Let
$$\xv(m)=[x_1^{N},\dots,x_m^{N}], \quad
\yv(m)=[y_1^{N},\dots,y_m^{N}],  \quad \text{and}\quad \zv(m)=[z_1^{N},\dots,z_m^{N}]$$
denote the input, the output at the intended receiver, and the output at the eavesdropper
after $m$ transmissions, respectively. For a given channel pair $(\hv,\gv)$, the average
error probability after the $m$ transmissions is defined as
\begin{equation}\label{pe}
    P_e(m|\hv) = \sum_{w \in \Wc} \Pr\left\{\phi\bigl(\Yv_{m}(w)\bigr) \neq w | w
    ~\text{sent},\hv\right\}\Pr(w),
\end{equation}
where $\phi\bigl(\Yv_{m}(w)\bigr)$ denotes the output of the decoder at the
legitimate receiver after $m$ transmissions.

The secrecy level after $m$ transmissions is given by
$$\frac{1}{{mN}}H(W|\Zv_{m}, \gv).$$
We say that perfect secrecy is achieved after $m$ transmissions if, for all
$\epsilon>0$, the equivocation rate satisfies
\begin{equation}\label{fequivoc}
    \frac{1}{mN}H(W|\Zv_{m}, \gv) \geq \frac{1}{mN}H(W)-\epsilon.
\end{equation}
We note that this definition implies that the perfect secrecy can
also be achieved after $j$ transmissions, for $j=1,\dots,m-1$.

Similar to the definition of good codes for a single-block transmission, we say
that a code $C$ of length $mN$ is \emph{good} for the $m$-block transmission
and a channel pair $(\hv,\gv)$ if $P_e(m|\hv) \leq \epsilon$ and the perfect
secrecy requirement (\ref{fequivoc}) can be achieved, for all $\epsilon>0$ and
sufficiently large $N$.

In particular, we consider the following two secure HARQ protocols based on
different mother codes and different combination techniques.

\subsubsection{Incremental Redundancy}

In the INR secure HARQ protocol, the mother code is a Wyner code of length
$MN$, i.e.,
$$C \in \Cc(R_0,R_s, MN).$$
In the first transmission, the transmitted coded symbols $\xv(1)=[x_1^{N}]$
form a codeword of a punctured Wyner code of length $N$,
$$C_1\in \Cc\left(MR_0, MR_s, N\right).$$
Similarly, after $m$ transmission, $m=1,\dots,M$, the (all)
transmitted coded symbols $\xv(m)=[x_1^{N},\dots,x_m^{N}]$ form a
codeword of a punctured Wyner code of length $mN$,
$$C_m\in \Cc\left(\frac{MR_0}{m}, \frac{MR_s}{m}, mN\right).$$
At the legitimate receiver and the eavesdropper, decoding and
equivocation calculation are attempted, respectively, based on the
punctured code $C_m$.

We note that the punctured codes $\{C_M,\,C_{M-1},\,\dots,\,C_1\}$
form a family of {\it rate-compatible} Wyner codes with the secrecy
rates
$$\left\{R_s,\,\frac{M}{M-1}R_s,\,\dots,\,MR_s\right\}.$$
Hence, we refer to this protocol as the INR protocol based on rate-compatible
Wyner codes.

\subsubsection{Repetition Time Diversity}

We also consider a simple time-diversity HARQ protocol based on the repetition
of a Wyner code. In this case, the mother code $C$ is a concatenated code
consisting of the Wyner code $C_1\in\Cc\left(MR_0, MR_s, N\right)$ as the outer
code and a simple repetition code of length $M$ as the inner code, i.e.,
\begin{equation}\label{rptcode}
    C=[\underbrace{C_1, C_1, \dots, C_1}_{M}].
\end{equation}
After each transmission, decoding and equivocation calculation are performed at
the receiver and the eavesdropper, respectively, based on maximal-ratio packet
combining.

\section{Secure Channel Set and Outage Events}\label{sec:coding}

In this section, we study the error performance and the secrecy level when a
mother Wyner code is transmitted over $M$ parallel channels. Results given in
this section form the basis for the performance analysis of secure HARQ
protocols.

For a given Wyner code, an important practical question is: under what channel
conditions will the communication be reliable and secure? In the following
theorem, we describe a {\it secure channel set} and demonstrate that there
exists a Wyner code sequence good for all channel pairs in this set.

\begin{theorem} \label{thm:Mp}
Let $\Pc$ denote the union of all channel pairs $(\hv,\gv)$
satisfying
\begin{align}
&           &  \frac{1}{M}\sum_{i=1}^{M}I(X;Y | h_i) &\ge R_0 &\label{M_secapA}\\
&\text{and} &  \frac{1}{M}\sum_{i=1}^{M}I(X;Z | g_i) &\le R_0-R_s, &
\label{M_secapB}
\end{align}
where $I(X;Y | h_i)$ and $I(X;Z | g_i)$ are single letter mutual information
characterizations of the channel (\ref{eq:chm}). There exists a Wyner code $C
\in \Cc(R_0, R_s, MN)$ good for all channel pairs $(\hv, \gv)\in \Pc$.
\end{theorem}

\begin{IEEEproof}
A proof of Theorem~\ref{thm:Mp} is provided in Appendix~\ref{app:lemma1}.
\end{IEEEproof}

In the system model described in Section~\ref{sec:model}, the transmitter does
not have any channel state information; that is, one cannot choose the code
based on a particular fading channel state. Hence, it is important to show that
there exists a Wyner code sequence good for all channel pairs in the secure
channel set $\Pc$.

To facilitate the formulation of outage-based throughput, we define
that an outage event occurs when the channel pair does not belong to
the secure channel set, i.e., $(\hv,\gv)\notin \Pc$. Specifically,
we distinguish two types of outage: \emph{connection outage}
\footnote{The main channel is viewed as a communication link. The
link is connected if a packet can be delivered to the intended
receiver successfully within the delay constraint (within $M$
transmissions), otherwise it is in the connection outage. The
connection outage probability defined in this paper is also referred
to as \emph{information outage probability} in \cite{Shamai:VT:94}.}
and \emph{secrecy outage}. In particular, we say that a connection
outage occurs if
\begin{align}
\frac{1}{M}\sum_{i=1}^{M}I(X;Y | h_i)  &<  R_0,
\end{align}
while we say that a secrecy outage occurs if
\begin{align}
\frac{1}{M}\sum_{i=1}^{M}I(X;Y | g_i)   & >  R_0-R_s.
\end{align}

Accordingly, we can evaluate both connection outage and secrecy
outage probabilities, which are the probabilities of each of the
outage events averaged over all possible fading states. In fact, the
connection outage probability can be interpreted as the limiting
error probability for large block length packets; the secrecy outage
probability can be regarded as an upper bound on the probability of
unsecured packets. Moreover, Theorem~\ref{thm:Mp} implies that the
connection outage probability and the secrecy outage probability are
not just average probabilities over a code ensemble, but they can be
achieved by a deterministic code sequence.

\section{Secure HARQ with Wyner Codes}\label{sec:sHARQ}

In this section, we evaluate the error performance and measure the
secrecy level during secure HARQ sessions.

A key part of an ARQ protocol is that decoding errors should be detected, so
that ACKs or NACKs can be generated accurately. A \emph{complete decoding
function} (e.g. maximum a posteriori probability decoding or maximum-likelihood
decoding) requires the encoder to add extra redundancy to the information bits,
which decreases the throughput slightly. The authors of \cite{Caire:IT:01} have
shown that error detection can be accomplished by using the built-in error
detection capability of suboptimal decoders.
\begin{lemma}\cite[Lemma~$3$]{Caire:IT:01}
For all $\epsilon>0$ and channel $\hv$, any code $C$ of length $MN$ satisfies
$$\Pr\left(\mathrm{undetected~error} | \hv, C \right) < \epsilon,$$ for
all sufficiently large $N$.
\end{lemma}
\begin{IEEEproof}
The proof follows similarly to that given in \cite{Caire:IT:01}.
\end{IEEEproof}

\subsection{Incremental Redundancy}

\begin{figure}[bt]
\includegraphics[width=0.55\linewidth]{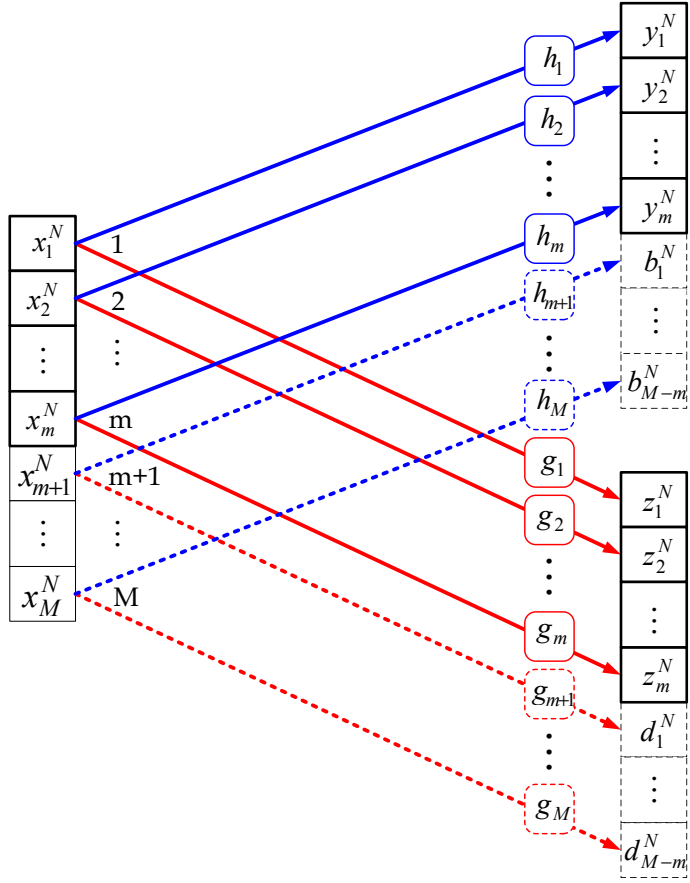}\\
  \centering
  \caption{$M$-parallel channel model for the INR protocol:
  the first $m$ punctured blocks are actually transmitted (solid lines);
  the remaining $M-m$ punctured blocks are assumed to be sent via $M-m$ dummy memoryless channels whose outputs are independent of the inputs (dashed lines).}
  \label{fig:dummy}
\end{figure}
To evaluate the performance of the INR protocol, we employ the following
$M$-parallel channel model. Let us focus on the decoding after $m$
transmissions, i.e., the coded blocks $\xv(m)=[x_1^{N},\dots,x_m^{N}]$ are
transmitted, $m=1,\dots, M$. As shown in Fig.~\ref{fig:dummy}, the block
$x_i^{N}$ experiences channel pair $(h_i,g_i)$, $i=1,\dots,m$. We assume that
each of the punctured blocks $[x_{m+1}^{N},\dots,x_{M}^{N}]$ is sent to a dummy
memoryless component channel whose output is independent of the input.

In this case, the mother codeword is transmitted over $M$ parallel channels. At
the legitimate receiver, the decoder combines the real signal
$\yv(m)=[y_1^{N},\dots,y_m^{N}]$ with $M-m$ dummy signal blocks
$[b_{1}^{N},\dots,b_{M-m}^{N}]$ to form
$$[y_1^{N},\dots,y_m^{N}, b_{1}^{N},\dots,b_{M-m}^{N}].$$
Similarly, the processed symbols at the eavesdropper are
$$[z_1^{N},\dots,z_m^{N}, d_{1}^{N},\dots,d_{M-m}^{N}],$$
where $[d_{1}^{N},\dots,d_{M-m}^{N}]$ are $M-m$ dummy signal blocks.
We note that the added dummy blocks do not affect either the
decoding at the legitimated receiver or the equivocation calculation
at the eavesdropper since they are independent of the confidential
message.

The codewords of the mother Wyner code $C$ are transmitted in at most $M$
transmissions during the secure HARQ session. By using the equivalent parallel
channel model, we can describe this secure HARQ problem as communication over
$M$ parallel wire-tap channels and, hence, establish the following theorem.

\begin{theorem} \label{thm:INR}
Consider the secure INR protocol based on rate compatible Wyner codes
$$\{C_M,C_{M-1},\dots,C_1\},$$
where
\begin{align*}
C_m\in \Cc\left(\frac{MR_0}{m}, \frac{MR_s}{m}, mN\right), \quad  m=1,\dots,M.
\end{align*}
Let $\Pc(m)$ denote the union of all channel pairs $(\hv,\gv)$
satisfying
\begin{align}
&           & \frac{1}{M}\sum_{i=1}^{m}I(X;Y | h_i) &\geq R_0, &\label{IRARQ_secapA}\\
& \text{and}& \frac{1}{M}\sum_{i=1}^{m}I(X;Z | g_i) &\leq R_0-R_s.
&\label{IRARQ_secapB}
\end{align}
Then, there exists a family of rate compatible Wyner codes
$\{C_M,C_{M-1},\dots,C_1\}$ such that $C_m$ is good for all channel pairs
$(\hv, \gv)\in \Pc(m)$, for $i=1,\dots,M$.
\end{theorem}
\begin{IEEEproof}
We provide a proof of Theorem~\ref{thm:INR} in Appendix
\ref{app:inr}.
\end{IEEEproof}

\subsection{Repetition Time Diversity}

In the RTD secure HARQ protocol, both the legitimate receiver and the
eavesdropper combine several noisy observations of the same packet based on
diversity techniques. The optimal receivers perform maximal-ratio combining
(MRC), which essentially transforms the vector channel pair $(\hv,\gv)$ into a
scalar channel pair $(\hat{h}(m),\hat{g}(m))$. Hence, after $m$ transmissions,
the equivalent channel model can be written as follows:
\begin{equation}\label{cobch}
  y(t) = \sqrt{\hat{h}(m)}x(t) + v(t)\quad\text{and}\quad
  z(t) = \sqrt{\hat{g}(m)}x(t) + u(t)
\end{equation}
for $t=1,\dots,N$, where $\hat{h}(m)=\sum_{i=1}^{m}h_i$ and
$\hat{g}(m)=\sum_{i=1}^{m}g_i$.

Let $\Lc(m)$ denote the union of all channel pairs $(\hv,\gv)$
satisfying
\begin{align}
&       &      I(X;Y | \hat{h}(m)) &\geq MR_0, &\label{RDTA}\\
& \text{and} & I(X;Z | \hat{g}(m)) &\leq M(R_0-R_s), &\label{RDTB}
\end{align}
where $I(X;Y | \hat{h}(m))$ and $I(X;Z | \hat{g}(m))$ are single letter mutual
information characterizations of the channel (\ref{cobch}). For a given
(finite) $M$, we have the following result for the RTD secure HARQ protocol.
\begin{corollary}
There exists a Wyner code $C_1\in\Cc\left(MR_0, MR_s, N\right)$ such that its
$m$-repeating code
$$C_m=[\underbrace{C_1, C_1, \dots, C_1}_{m}]$$
is good for all channel pairs $(\hv, \gv)\in \Lc(m)$, for $m=1, \dots, M$.
\end{corollary}
\begin{IEEEproof}
The proof follows directly from Theorem~\ref{thm:Mp} by setting
$M=1$.
\end{IEEEproof}

\section{Secrecy Throughput of HARQ Protocols}\label{sec:secthru}

In this section, we study the achievable secrecy throughput for HARQ protocols.
We focus on Rayleigh independent block fading channels for illustration; other
types of block fading channels can be studied in a similar way.

We note that the optimal input distribution of the channel (\ref{eq:chm}) is
not known in general when the transmitter has no CSI. For the sake of
mathematical tractability, we consider Gaussian inputs. For INR, the mutual
information $I_{XY}^{[\rm INR]}(m)$ and $I_{XZ}^{[\rm INR]}(m)$ can be written
as
\begin{align}
  \nonumber
&  & I_{XY}^{[\rm INR]}(m) & = \frac{1}{2M} \sum_{i=1}^{m} \log_{2}\left(1+\lambda_i\right)&\\
&\text{and}& I_{XZ}^{[\rm INR]}(m) & = \frac{1}{2M} \sum_{i=1}^{m}
\log_{2}\left(1+\nu_i\right), & \label{gaussianMI}
\end{align}
where
\begin{equation}\label{snrs}
\lambda_i=h_i\bar{P} \quad \text{and} \quad \nu_i=g_i \bar{P},\quad
i=1,\dots,M,
\end{equation}
are the signal-to-noise ratios (SNRs) at the legitimate receiver and the
eavesdropper, respectively, during transmission~$i$. For RTD, we can express
the mutual information quantities $I_{XY}^{[\rm RTD]}(m)$ and $I_{XZ}^{[\rm
RTD]}(m)$ as
\begin{align}\label{reptMI}
  \nonumber
&   &  I_{XY}^{[\rm RTD]}(m) & = \frac{1}{2M}
  \log_{2}\left(1+\sum_{i=1}^{m}\lambda_i\right)\\
&\text{and}&  I_{XZ}^{[\rm RTD]}(m) & = \frac{1}{2M}
  \log_{2}\left(1+\sum_{i=1}^{m}\nu_i\right).&
\end{align}
Although we consider only Gaussian signaling here, the results in
Section~\ref{sec:sHARQ} can be applied to other input distributions, for
example, discrete signaling under modulation constraints.

Let $\Mc$ denote the number of transmissions within a HARQ session. Given a
distribution of the main channel SNR $\lambda$, for both INR and RTD protocols,
the probability mass function of $\Mc$ can be expressed as
\begin{align}
&   & p[\Mc=m] &= \Pr \left\{I_{XY}(m-1) < R_0 ~\text{and}~ I_{XY}(m) \geq R_0\right\}   &\notag\\
&    & &= \Pr \left\{I_{XY}(m-1) < R_0\right\}
      - \Pr\left\{I_{XY}(m) < R_0\right\},\quad m=1,\dots,M-1, & \notag\\
& \text{and}& p[\Mc=M]&=\Pr\left\{I_{XY}(M-1) < R_0\right\}, &\label{eq:pmf-m}
\end{align}
where $I_{XY}(m)$ and $I_{XZ}(m)$ are chosen either from
(\ref{reptMI}) or from (\ref{gaussianMI}) corresponding to a
specific HARQ protocol. Let $P_e$ denote the connection outage
probability, and $P_s$ denote the secrecy outage probability. The
definition in (\ref{eq:pmf-m}) implies that $P_e$ and $P_s$ can be
written as follows:
\begin{align}
&   &         P_e &= \Pr\left\{I_{XY}(M) <
R_0\right\},&\label{eq:def-pe} \\
&\text{and}&   P_s &=\sum_{m=1}^{M}p[m]\Pr\left\{I_{XZ}(m)
> R_0 - R_s\right\}. & \label{eq:def-ps}
\end{align}


Now, we study the secrecy throughput based on $P_e$ and $P_s$. We first
consider a target secrecy outage probability $\xi_s$; that is, at least a
fraction $1- \xi_s$ of the confidential message bits sent by the transmitter
are kept completely secret. Under this constraint, the secrecy throughput
$\eta$, measured in bits per second per hertz, is defined to be the average
number of bits decoded at the legitimate receiver,
\begin{equation}\label{eta}
    \eta=\lim_{t\rightarrow \infty}\frac{a(t)}{tN},
\end{equation}
where again $N$ is the number of symbols in each block and $a(t)$ is the number
of information bits successfully decoded by the intended receiver up to time
slot $t$ (when a total of $tN$ blocks are sent). The event that the transmitter
stops sending the current codeword is recognized to be a \emph{recurrent
event}\cite{Zorzi:TC:96}. A random \emph{reward} $\mathcal{R}$ is associated
with the occurrence of the recurrent event. In particular, $\mathcal{R}=MR_s$
bits/symbol if transmission stops because of successful decoding, and
$\mathcal{R}=0$ bits/symbol if it stops because successful decoding has not
occurred after $M$ transmissions. By applying the renewal-reward theorem
\cite{Caire:IT:01,Zorzi:TC:96}, we obtain the secrecy throughput as
\begin{equation}\label{eta_arq}
    \eta(R_0,R_s)=\frac{\E[\mathcal{R}]}{\E[\Mc]}=\frac{MR_s}{\E[\Mc]}(1-P_e),
\end{equation}
where $\E[\Mc]$ is the expected number of transmissions in order to complete a
codeword transmission, i.e.,
\begin{align}
\E[\Mc]&=\sum_{m=1}^{M}mp[\Mc=m]\notag\\
     &=1+\sum_{m=1}^{M}\Pr\left\{I_{XY}(m) < R_0\right\}.\label{Em}
\end{align}

We can properly choose the mother code parameters ($R_0$ and $R_s$)
to obtain the maximum throughput while satisfying $\xi_s$-secrecy
requirement. Hence, we consider the following problem
\begin{align}\label{eq:opt}
\max_{R_0, R_s} \quad &\eta(R_0,R_s)\\
\text{s.t.}\qquad  &P_s \leq \xi_s.
\nonumber
\end{align}
The optimization problem (\ref{eq:opt}) imposes a probabilistic service
requirement in terms of confidentiality; that is, the service quality is
acceptable as long as the probability of the secrecy outage is less than
$\xi_s$, a parameter indicating the outage tolerance of the application. Note
that $P_s$ is a decreasing function of $R_s$, and $\eta$ is linearly
proportional to $R_s$. Hence, we can solve the optimization problem
(\ref{eq:opt}) in the following two steps: first, for given $M$, $R_0$, and
$\xi_s$, we find the maximum value $R_s^*(R_0)$; next, we obtain the optimum
$R_0^*$, which maximizes the secrecy throughput $\eta(R_0,R_s^*(R_0))$.


On the other hand, reliability is another important quality of
service parameter. To achieve both the connection outage target
$\xi_e$ and the secrecy outage target $\xi_s$, we consider the
following problem
\begin{align}\label{eq:opt2}
&\max_{R_0, R_s}~~~~ \eta(R_0,R_s)\\
&~~\text{s.t.} \qquad ~~P_s \leq \xi_s, ~~P_e \leq \xi_e. \nonumber
\end{align}
In addition to the service requirement of confidentiality, problem
(\ref{eq:opt2}) also imposes a probabilistic service requirement on
the connection outage, i.e., at least a fraction $1-\xi_s$ of HARQ
sessions are successful. The connection outage constraint ensures
that, at the expense of possibly lower average throughput, the delay
constraint (that a packet can be delivered within $M$ transmissions)
is satisfied $1-\xi_s$ of the time, hence enabling applications
which trade average rate for decoding delay like voice communication
systems, e.g., CDMA2000 \cite{CDMA2000}. A similar constraint has
been considered in \cite{LuoJ:IT:05} in terms of {\it service
outage} for parallel fading channels.

To evaluate $p[m]$, $P_e$ and $P_s$, we need the cumulative
distribution functions (CDFs) of $I_{XY}(m)$ and $I_{XZ}(m)$. For
the RTD protocol, we can use the fact that $\sum_{i=1}^{m}\lambda_i$
and $\sum_{i=1}^{m}\nu_i$ are gamma distributed to express the CDFs
of $I_{XY}^{[\rm RTD]}(m)$ and $I_{XZ}^{[\rm RTD]}(m)$ in terms of
incomplete gamma functions. In the case of the INR protocol, the
distributions of $I_{XY}^{[\rm INR]}(m)$ and $I_{XZ}^{[\rm INR]}(m)$
cannot be written in a closed form. Hence, we resort to Monte-Carlo
simulation in order to obtain empirical CDFs. Note that Monte Carlo
simulation is needed only to estimate empirical CDFs, while
$(R_0^*,R_s^*)$ is found numerically by a (non-random) search.

\section{Asymptotic Analysis} \label{sec:limit}

In general, the secrecy throughput of the INR protocol is difficult to
calculate since there is no closed form available for $\Pr\{I_{XY}(m) < R_0\}$.
In this section, we consider the asymptotic secrecy throughput, which does have
a closed form.

We are interested in asymptotic results as $M$ increases without bound. Note
that this asymptote corresponds to a delay-unconstrained system. In this case,
secure HARQ protocols yield zero packet loss probability, i.e., the
transmission of a codeword ends only when it is correctly decoded. As a result,
the problems (\ref{eq:opt}) and (\ref{eq:opt2}) yield the same throughput,
which can be obtained from (\ref{eta_arq}) as follows:
\begin{equation}\label{eta_asmp0}
    \eta(R_0,R_s) = \frac{MR_s}{\E[\Mc]}=\frac{MR_s}{1+\sum_{m=1}^{M}\Pr\left\{I_{XY}(m) <
    R_0\right\}}.
\end{equation}

Let us consider how to choose a mother Wyner code for the INR protocol in order
to meet reliability and confidentiality constraints when $M$ is large. Let
$\lambda$ and $\nu$ denote the instantaneous SNRs at the legitimate receiver
and the eavesdropper, respectively.

\begin{lemma} \label{lem:large}
Consider an INR secure HARQ protocol with the mother Wyner code $C\in \Cc(R_0,
R_s, MN).$ Then
\begin{align}
\lim_{M \rightarrow \infty} P_e^{[\rm INR]} = 0 \quad \text{and} \quad \lim_{M
\rightarrow \infty} P_s^{[\rm INR]} =0, \label{eq:sec-re}
\end{align}
if and only if
\begin{align}
&   &          R_0 &  \le \frac{1}{2} \E[\log_2(1+\lambda)]& \notag\\
&\text{and} &  R_0 - R_s &\ge
R_0\frac{\E[\log_2(1+\nu)]}{\E[\log_2(1+\lambda)]}, & \label{eq:cond1}
\end{align}
where the expectations are over $\lambda$ and/or $\nu$. Furthermore, if
(\ref{eq:cond1}) does not hold, then
\begin{align}
\text{either} \quad \lim_{M \rightarrow \infty} P_e^{[\rm INR]} = 1
\quad \text{or} \quad \lim_{M \rightarrow \infty} P_s^{[\rm INR]}
=1. \label{eq:sec-re1}
\end{align}
\end{lemma}
\begin{IEEEproof}
A proof of Lemma~\ref{lem:large} is given in Appendix
\ref{app:limitlm2}.
\end{IEEEproof}

For comparison, we consider the situation in which the Wyner code $C$ is
transmitted over $M$-block fading channel without using the HARQ protocol. We
refer to this case as the $M$-fading-block (MFB) coding scheme.
Theorem~\ref{thm:Mp} implies that, by using the MFB scheme, the requirement
(\ref{eq:sec-re}) can be achieved if and only if
\begin{align}
&            &  R_0 &\le \frac{1}{2}\E[\log_2(1+\lambda)]& \notag \\
& \text{and} & R_0 - R_s &\ge \frac{1}{2}\E[\log_2(1+\nu)]. \label{eq:cond2}
\end{align}
We note that the condition (\ref{eq:cond1}) for the INR protocol is
weaker than the condition (\ref{eq:cond2}) for the MFB scheme. In
other words, the INR scheme can achieve the confidentiality and
reliability requirements more easily than can the MFB coding scheme
by using the same Wyner code. This result illustrates the benefit of
the INR secure HARQ protocol.

Based on Lemma~\ref{lem:large}, we have the following asymptotic result
concerning the achievable throughput for secure HARQ protocols.
\begin{theorem} \label{thm:lim}
We consider the secure HARQ protocols over a block-fading wire-tap channel. If
the secrecy information rate $R_0$ satisfies
\begin{align}
\lim_{M\rightarrow\infty} \frac{1}{MR_s}=0,
\end{align}
then the secrecy throughput of RTD and INR protocols can be written as follows:
\begin{eqnarray*}
\lim_{M\rightarrow \infty}\; \max_{R_0,R_s}\eta(R_0,R_s) = \left\{
\begin{array}{rl}
  \quad 0 \qquad \qquad \qquad \qquad \qquad &\mbox{ RTD} \\
  (1/2) \E \left[\log_2(1+\lambda) - \log_2(1+\nu)\right] &\mbox{ INR}
       \end{array} \right. ,
\end{eqnarray*}
where $\lambda$ and $\nu$ are the instantaneous SNRs at the
legitimate receiver and the eavesdropper, respectively.
\end{theorem}
\begin{IEEEproof}
We provide a proof in Appendix~\ref{app:RTD}.
\end{IEEEproof}
We note that the RTD protocol involves suboptimal coding schemes,
for which $\E[\Mc]$ grows faster than $MR_s$ in (\ref{eta_asmp0}).
Hence, the limiting secrecy throughput $\eta$ is zero.
Theorem~\ref{thm:lim} again asserts the benefit of INR over RTD.

\section{Numerical Results}
\label{sec:sac}

In our numerical examples, we consider Rayleigh block fading, i.e. the main
channel instantaneous SNR $\lambda$ has the probability density function (PDF)
$f(\lambda)=(1/\bar{\lambda})e^{-\lambda/\bar{\lambda}}$,  and the eavesdropper
channel instantaneous SNR $\nu$ has the PDF
$f(\nu)=(1/\bar{\nu})e^{-\nu/\bar{\nu}}$, where $\bar{\lambda}$ and $\bar{\nu}$
are the average SNRs of the main and eavesdropper channels, respectively.

\begin{figure}[bt]
  \includegraphics[width=0.65\linewidth]{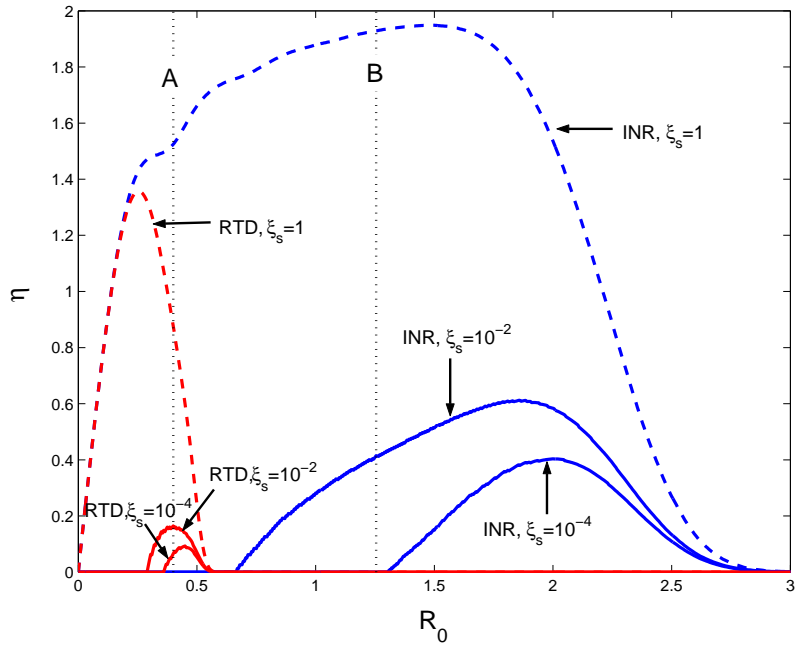}\\
  \centering
  \caption{Secrecy throughput $\eta$ versus the main channel code rate $R_0$ under different secrecy requirements $\xi_s$,
 where the main channel average SNR is 15dB, the eavesdropper channel
  average SNR is $5$dB, and the maximum number of transmissions is $M=8$.}\label{fig:thru_R0}
\end{figure}

\begin{figure}[bt]
  \centerline{\includegraphics[width=0.65\linewidth]{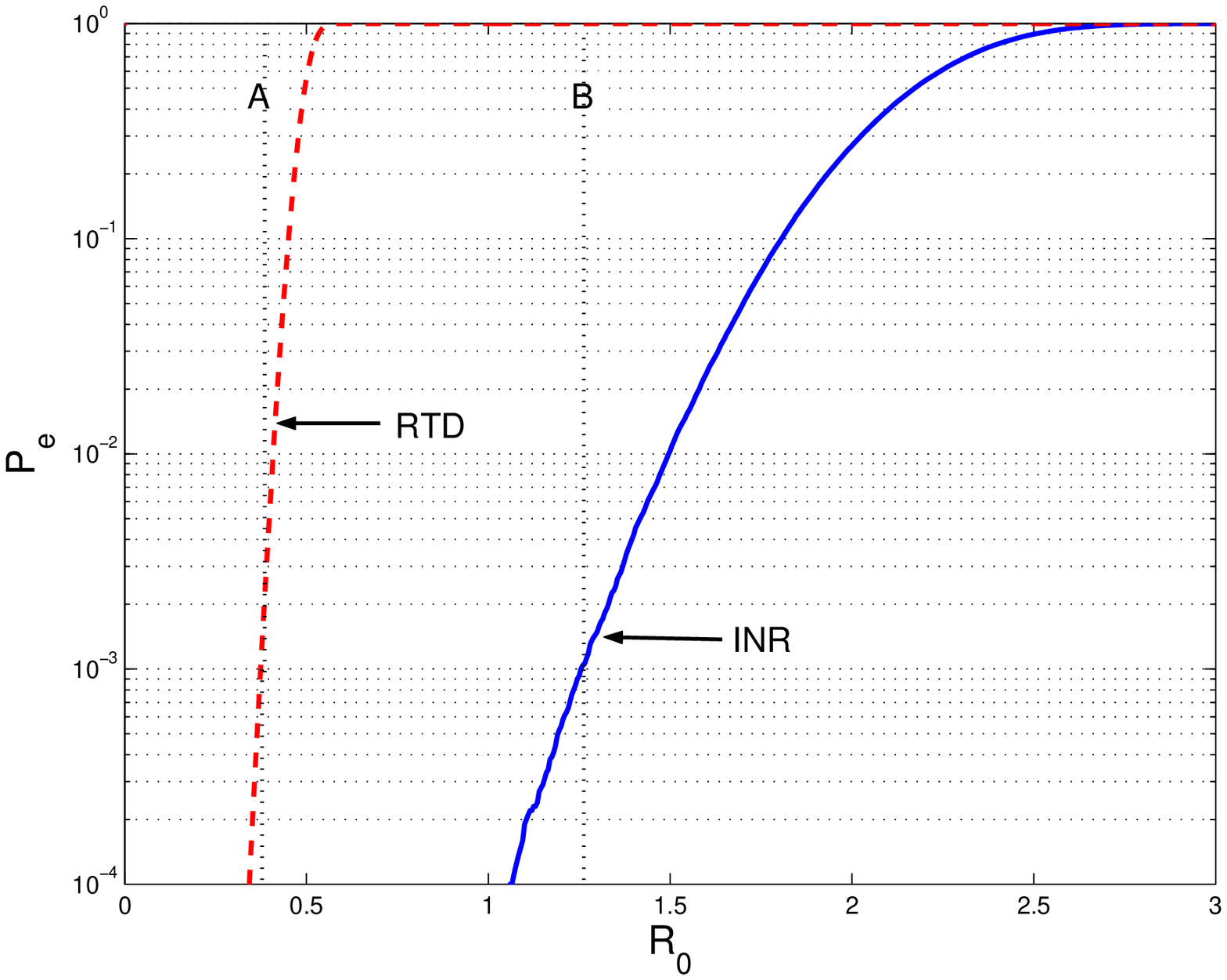}}
   \caption{Connection outage probability $P_e$ versus the main channel code rate $R_0$, where the main channel average SNR is $15$dB, the eavesdropper
channel average SNR is $5$dB, and $M=8$.}\label{pe_R0}
\end{figure}

To illustrate how the secrecy throughput $\eta$ is related to the
choice of $R_0$ (and $R_s$), we give a numerical example of $\eta$
versus $R_0$ in Fig.~\ref{fig:thru_R0}, in which the parameter
settings are as follows: the main channel average SNR
$\bar{\lambda}$ is $15$dB, the eavesdropper channel average SNR
$\bar{\nu}$ is $5$dB, the maximum number of transmissions $M$ is
$8$. (We observe that similar results are obtained by using other
parameter settings.) For each $R_0$, we obtain the maximum
$R_s^*(R_0)$ that meets the secrecy constraint $\xi_s=1, 10^{-2}$ or
$10^{-4}$, respectively. When there is no secrecy constraint
($\xi_s=1$), due to the sub-optimality of the RTD scheme, the RTD
curve is uniformly below the INR curve. This does not happen when
there is a secrecy constraint. The reason is that INR not only
favors the information transmission to the intended receiver, but
also benefits the eavesdropping by the eavesdropper. Hence, INR
needs to sacrifice a larger portion of the main channel code rate
than RTD in order to keep the eavesdropper ignorant of the
confidential messages. This is reflected in Fig. \ref{fig:thru_R0}
that a larger $R_0$ has to be chosen for INR (than RTD) in order to
obtain a positive secrecy throughput.

It is clear from Fig.~\ref{fig:thru_R0} that there exists a unique
$R_0^*$ (and therefore $R_s^*(R_0^*)$) to maximize $\eta$ for each
parameter setting. For all secrecy constraints ($\xi_s=1, 10^{-2}$
or $10^{-4}$), if the best $R_0^*$ and $R_s^*(R_0^*)$ are chosen for
each scheme accordingly, INR yields higher secrecy throughput than
RTD does, which shows the benefit of INR over RTD.

According to (\ref{eq:def-pe}), the choice of $R_0$ decides the
reliability performance.  This is shown in Fig. \ref{pe_R0}, where
we plot the connection outage probability $P_e$ versus the value of
$R_0$. For both INR and RTD, $P_e$ increases with the value of
$R_0$. Note that a more strict secrecy constraint requires a larger
$R_0^*$ (as shown in Fig. \ref{fig:thru_R0}), which however causes
the degradation of the reliability performance. We can see that
there exists a tradeoff between secrecy and reliability.

Given a strict connection outage constraint $P_e<\xi_e$, the choice
of $R_0^*$ (and $R_s^*(R_0^*)$) might not be feasible. For instance,
in order to obtain $P_e< 10^{-3}$, we need to choose
$R_0^{[\mathrm{RTD}]} \leq 0.38$ and $R_0^{[\mathrm{INR}]} \leq
1.25$ (marked with `A' and `B' respectively in Fig.
\ref{fig:thru_R0} and Fig. \ref{pe_R0}). Specifically, for a
connection outage constraint $P_e<10^{-3}$, $R_0^*$ is not feasible
for INR when $\xi_s=10^{-2}$, and $R_0^*$ is not feasible for both
INR and RTD when $\xi_s=10^{-4}$ in Fig. \ref{fig:thru_R0}. Note
that for the case of $\xi_s=10^{-4}$ (and $\xi_e=10^{-3}$), positive
secrecy throughput cannot be obtained for INR, but can be obtained
for RTD. This implies that RTD might outperform INR, when we have
strict secrecy and connection outage constraints. This is a
surprising result in the view of the well-known HARQ performance
when there is no secrecy constraint, where INR always outperforms
RTD \cite{Caire:IT:01}.

\begin{figure}[bt]
  \centerline{\includegraphics[width=0.65\linewidth]{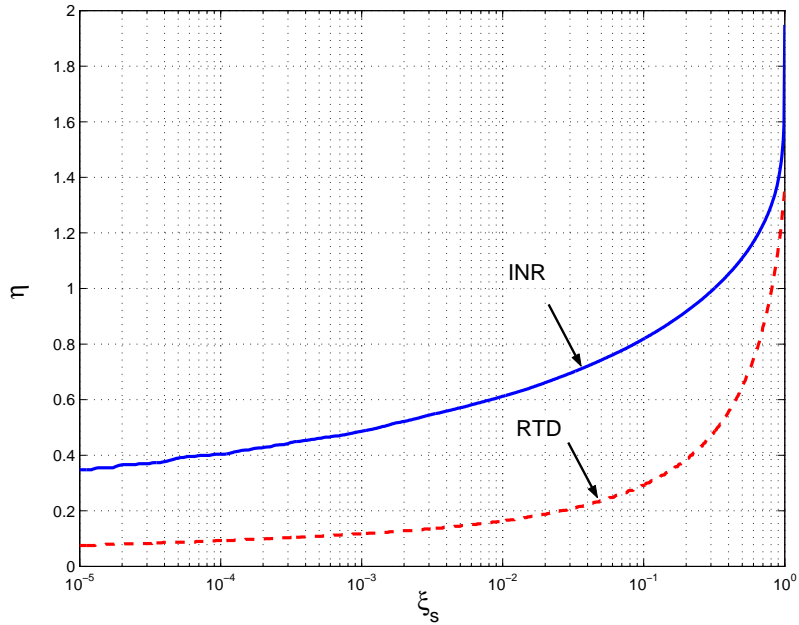}}
   \caption{Throughput $\eta$ versus target secrecy outage probability $\xi_s$,
   when the main channel average SNR is $15$dB, the eavesdropper channel
  average SNR is $5$dB, and $M=8$.}\label{ps_thru-npe}
\end{figure}

\begin{figure}[hbt]
  \centerline{\includegraphics[width=0.65\linewidth]{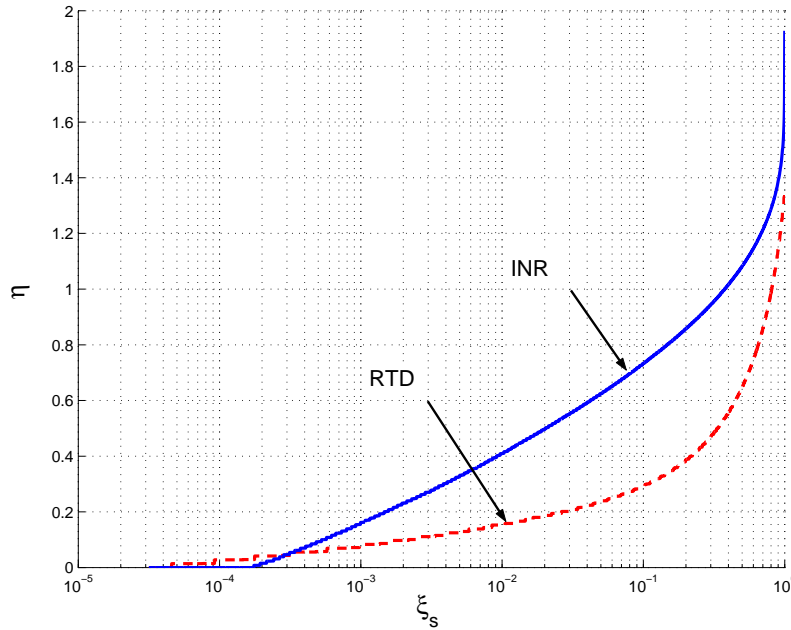}}
  \caption{Throughput $\eta$ versus target secrecy outage probability $\xi_s$ under
  connection outage probability $\xi_e=10^{-3}$,
  when the main channel average SNR is $15$dB, the eavesdropper channel
  average SNR is $5$dB, and $M=8$.}\label{ps_thru}
\end{figure}

In Fig. \ref{ps_thru-npe} and Fig. \ref{ps_thru}, we show the
secrecy throughput $\eta$ under different target secrecy outage
probabilities $\xi_s$. There is no connection outage requirement in
Fig. \ref{ps_thru-npe}. There is an additional connection outage
requirement of $p_e \leq \xi_e = 10^{-3}$ in Fig. \ref{ps_thru}. The
parameter settings are $\bar{\lambda}=15$dB, $\bar{\nu}=5$dB and
$M=8$. We can see that small secrecy outage probability can be
achieved when the throughput is small for both protocols. The INR
protocol outperforms the RTD protocol uniformly when there is no
connection outage requirement. However, when there is a strict
connection outage requirement, the RTD protocol outperforms the INR
protocol when $\xi_s$ is small (e.g., $\xi_s \leq 10^{-4}$).

\begin{figure}[hbt]
  \centerline{\includegraphics[width=0.65\linewidth]{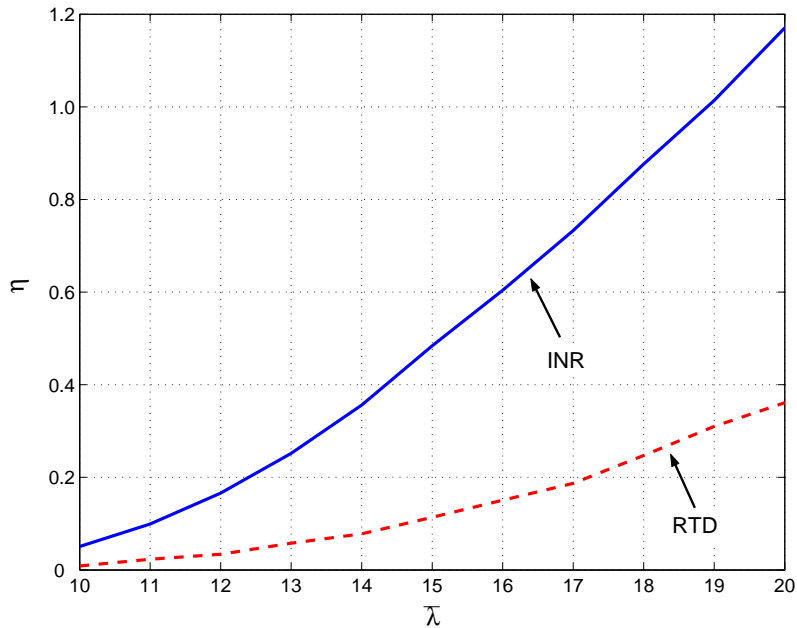}}
  \caption{Throughput $\eta$ versus main channel average SNR $\bar{\lambda}$
  under a target secrecy outage probability $\xi_s=10^{-3}$,
  when the eavesdropper channel average SNR is $5$dB and $M=8$.}\label{thru_snr}
\end{figure}
\begin{figure}[hbt]
  \centerline{\includegraphics[width=0.65\linewidth]{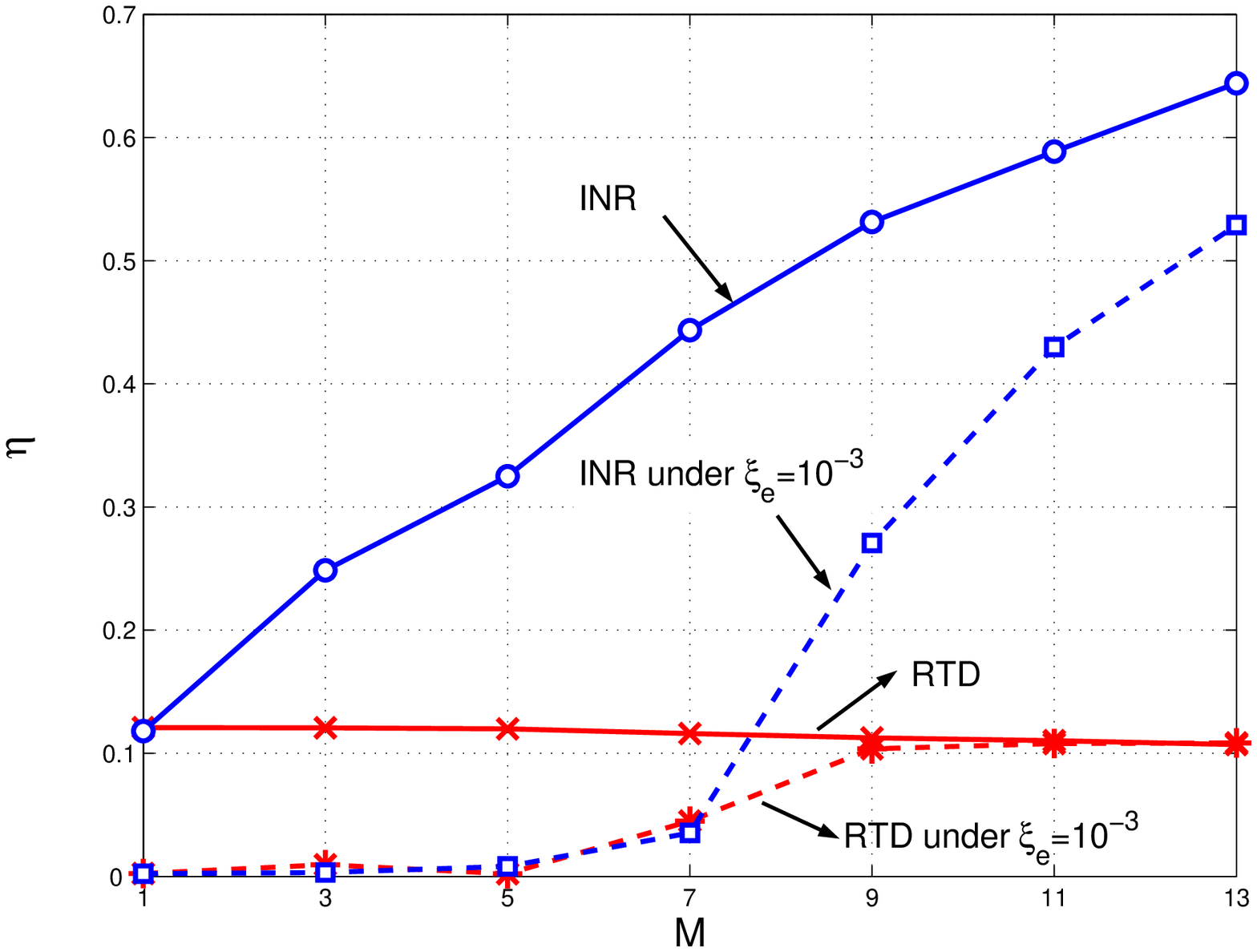}}
  \caption{Throughput $\eta$ versus the maximum number of transmissions $M$
  under a target secrecy outage probability $\xi_s=10^{-3}$, when the main and
  eavesdropper channel average SNRs are $15$dB and $5$dB, respectively.}\label{thru_M}
\end{figure}

Fig. \ref{thru_snr} illustrates the relationship between the secrecy
throughput $\eta$ and the main channel average SNR $\bar{\lambda}$
when there is a target secrecy outage probability $\xi_s=10^{-3}$
and no connection outage requirement. The average SNR of the
eavesdropper channel is fixed to be $5$dB. We find that the INR
protocol outperforms the RTD protocol significantly, especially when
the main channel SNR is large.

In Fig. \ref{thru_M}, we show the secrecy throughput $\eta$ versus
the maximum number of transmissions $M$. Comparing with the secrecy
throughput without the connection outage constraint, the secrecy
throughput with a connection outage constraint ($P_e \leq 10^{-3}$)
suffers some loss when $M$ is small due to insufficient diversity.
Both secrecy throughputs converge when sufficient diversity can be
obtained as $M$ increases. In particular, when $M \rightarrow
\infty$, both throughputs are the same and are given by
(\ref{eta_asmp0}) in the asymptotic analysis. For INR, the secrecy
throughput $\eta^{[\mathrm{INR}]}$ increases monotonically with $M$.
For RTD, $\eta^{[\mathrm{RTD}]}$ decreases with $M$ due to its
strongly suboptimal coding scheme. This concurs with the asymptotic
analysis that, when $M \rightarrow \infty$, a constant (nonzero)
secrecy throughput ($0.5*\E \left[\log_2(1+\lambda) -
\log_2(1+\nu)\right]= 1.31$ according to Theorem~ $3$) can be
achieved for INR, while zero throughput can be obtained for RTD.

\section{Conclusions and Future Directions}\label{sec:conclusions}

In this paper, we have studied secure packet communication over
frequency-flat block-fading Gaussian channels, based on secure HARQ
protocols with the joint consideration of channel coding, secrecy
coding and retransmission protocols. From an information theoretic
point of view, we have considered two secure HARQ protocols: a
repetition time diversity scheme with maximal-ratio combining (RTD),
and an incremental redundancy scheme based on rate-compatible Wyner
secrecy codes (INR). We have proved the existence of good Wyner code
sequences, which ensure that the legitimate receiver can decode the
message and the eavesdropper can be kept ignorant of it for an HARQ
session under certain channel realizations.

To facilitate the formulation of the outage-based throughput, we
have defined two types of outage: connection outage and secrecy
outage. The outage probabilities, more specifically, the connection
and secrecy outage probabilities have been used to characterize the
tradeoff between the reliability of the legitimate communication
link and the confidentiality with respect to the eavesdropper's
link. We have evaluated the achievable throughput of RTD and INR
protocols under probabilistic requirements (constraints) on secrecy
outage and/or connection outage, and have illustrated the benefits
of HARQ schemes to information secrecy through some numerical
results and an asymptotic analysis.

In general, INR can achieve a significantly larger throughput than
RTD, which concurs with the results not involving secrecy that
mutual-information accumulation (INR) is a more effective approach
than SNR-accumulation (RTD). However, when one is forced to ensure
small connection outage for the main channel even when it is bad,
one is forced to reduce the main channel code rate. The INR scheme,
having a larger coding gain (to both the intended receiver and the
eavesdropper), needs to sacrifice a larger portion of the main
channel code rate (i.e., requires a larger secrecy gap) in order to
satisfy the secrecy requirement. Hence when the main channel code
rate is bounded due to the connection outage constraint, the
achievable secrecy throughput of INR may be smaller than that of
RTD.

We conclude this work by pointing out some future research
directions.

First, as pointed out in \cite{Ghanim:CISS:06}, many practical
encoders are separated from the modulator and therefore the
performance of HARQ protocols is impacted by modulation constraints.
Although we have assumed Gaussian signaling, it is possible and also
meaningful to extend the analysis to take discrete signaling into
account.

In our analysis, we have assumed random coding and typical set
decoding. Future work should consider practical coding and decoding
schemes for secure HARQ protocols. Existing work on the practical
secrecy code design includes coset coding \cite{Ozarow:BSTJ:84},
low-density parity check (LDPC) code design \cite{Thangaraj:IT:07},
and nested codes \cite{Liu:ITW:07}. The design of practical rate
compatible secrecy codes for Gaussian channels remains a challenging
problem.

\appendices

\section{Proof of Theorem~\ref{thm:Mp}} \label{app:lemma1}

For convenience, let $\pv \triangleq (\hv,\gv)$ and $\Pc_{\ast}$
denote the set of channel pairs $(\hv,\gv)$ so that
\begin{align}
&            & \frac{1}{M}\sum_{i=1}^{M}I(X;Y | h_i)  &=  R_0 + \delta \label{Mfad_secap1}&\\
& \text{and} & \frac{1}{M}\sum_{i=1}^{M}I(X;Z | g_i)  &=  R_0 - R_s + \delta,
\label{Mfad_secap2}&
\end{align}
where $\delta>0$ is arbitrarily small. It is clear that $\Pc_{\ast}
\subseteq \Pc$ when $\delta \rightarrow 0$.

In order to prove Theorem~\ref{thm:Mp}, we first consider the
following lemma.
\begin{lemmaA1}
There exists a code $C \in \Cc(R_0, R_s, MN)$ that is good for any channel pair
$\pv \in \Pc_{\ast}$.
\end{lemmaA1}

\subsection{Proof of Lemma~$A.1$}
\begin{IEEEproof}
Following standard continuity arguments \cite{Cover:91}, we
consider a quantization of the input and output of the channel
(\ref{eq:chm}) and work on the resulting discrete channel. Given a
channel pair $\pv=(\hv,\gv)$, on every fading block $i \in [1,M]$,
the channel is time-invariant and memoryless. Let $\xv$ denote the
input, and let $\yv$ and $\zv$ denote the outputs at the
legitimate receiver and the eavesdropper, respectively. From the
weak law of large numbers, we have the following limits in
probability:
\begin{align*}
& & \lim_{N \rightarrow \infty} \frac{1}{N}\log_{2}\Pr(\xv)&=-MH(X), &\\
& & \lim_{N \rightarrow \infty} \frac{1}{N}\log_{2}\Pr(\yv)&=-\sum_{i=1}^{M}H(Y|h_i),& \\
& & \lim_{N \rightarrow \infty} \frac{1}{N}\log_{2}\Pr(\zv)&=-\sum_{i=1}^{M}H(Z|g_i),& \\
& & \lim_{N \rightarrow \infty} \frac{1}{N}\log_{2}\Pr(\xv,\yv)&=-\sum_{i=1}^{M}H(X,Y|h_i),& \\
& \text{and}& \lim_{N \rightarrow \infty}
\frac{1}{N}\log_{2}\Pr(\xv,\zv)&=-\sum_{i=1}^{M}H(X,Z|g_i), &
\end{align*}
where $H(X)$ is the input entropy per letter; $H(Y|h_i)$ and $H(Z|g_i)$ are the
output entropy per letter at the intended receiver and the eavesdropper,
respectively, in block $i=1,\dots,M$; and $H(X,Y|h_i)$ and $H(X,Z|g_i)$ are the
joint entropies per letter in block $i$. Define the typical set
$T_{\epsilon}^{N}$ as the set of all sequences $(\xv,\yv,\zv)$ for which the
above sample means are within $\epsilon$ of their limits.

The random coding ensemble $\Cc=\Cc(R_0, R_s, MN)$ is constructed by generating
$2^{NMR_0}$ codewords $\xv(w,v)$, where $w=1,2,\dots, 2^{NMR_s}$ and
$v=1,2,\dots,2^{NM(R_0-R_s)}$, by choosing the $(MN)2^{NMR_0}$ symbols
independently at random. Given $w \in \Wc=\{1, 2, \dots, 2^{NM{R_s}}\}$, the
encoder randomly and uniformly selects a $v$ from
$\{1,2,\dots,2^{NM(R_0-R_s)}\}$ and transmits $\xv(w,v)$.

\subsubsection{Error Analysis}

Given a message $w \in \Wc$, the legitimate receiver declares that $\xv$ was
transmitted, if $\xv$ is the only codeword that is jointly typical with $\yv$.
An error is declared if either $\xv$ is not jointly typical with $\yv$, or
there is another codeword $\tilde{\xv}$ jointly typical with $\yv$. Let us
denote this type of error as $\mathcal{E}_1$. By following the same steps in
\cite[Theorem~$8.7.1$]{Cover:91}, we obtain that $\E_{C \in
\Cc}[\Pr(\mathcal{E}_1 | \pv, C)]$, the probability of error $\mathcal{E}_1$
averaged over the code ensemble $\Cc$ is
\begin{eqnarray*}
\E_{C \in \Cc}[\Pr(\mathcal{E}_1 | \pv, C)]
     &\leq& \E\left\{\Pr\left[(\xv, \yv) \notin T_{\epsilon}^{N}(P_{XY})\right] + \sum_{\tilde{\xv} \neq \xv}\Pr\left[(\tilde{\xv}, \yv) \in T_{\epsilon}^{N}(P_{XY})\right]\right\}\\
   &\leq& \epsilon + (2^{NMR_0}-1)\E\left\{\Pr\left[(\tilde{\xv}, \yv) \in T_{\epsilon}^{N}(P_{XY})\right]\right\}\\
   &=& \epsilon + (2^{NMR_0}-1)2^{-N\left[\sum_{i=1}^{M}I(X;Y | h_i)-\epsilon\right]} \\
   &\leq& \epsilon + 2^{-N(\delta-\epsilon)}.
 \end{eqnarray*}
By choosing $\delta > \epsilon$, we have
\begin{equation}\label{ebound1}
    \E_{C \in \Cc}[\Pr(\mathcal{E}_1| \pv, C]  \leq \epsilon_1,
\end{equation}
for every channel pair $\pv \in \Pc_{\ast}$ as the codeword length $N$ is
sufficiently large, where $\epsilon_1=\epsilon + 2^{-N(\delta-\epsilon)}$.

Let $B(w)$ denote the set of codewords corresponding to message $w \in \Wc$
(bin $w$). Suppose that the eavesdropper gets to know $w$ a priori, based on
which it tries to determine which codeword was sent. The eavesdropper declares
that $\xv$ was sent, if $\xv$ is the only codeword in $B(w)$ that is jointly
typical with $\zv$. An error is declared if either $\xv$ is not jointly typical
with $\zv$, or there is another codeword $\tilde{\xv}$ in $B(w)$ jointly
typical with $\zv$. Denoting this type of error as $\mathcal{E}_2$, we obtain
that $\E_{C \in \Cc}[\Pr(\mathcal{E}_2 | \pv, C)]$, the average probability of
error averaged over the code ensemble $\Cc$ is
\begin{align*}
  \E_{C \in \Cc}[\Pr(\mathcal{E}_2 | \pv, C)]
   &\le \E\left\{\Pr\left[(\xv, \zv) \notin T_{\epsilon}^{N}(P_{XY})\right] + \sum_{\tilde{\xv} \neq \xv}\Pr\left[(\tilde{\xv}, \zv) \in T_{\epsilon}^{N}(P_{XZ}),\tilde{\xv} \in B(w)\right]\right\}\\
   &\le \epsilon + (2^{NMR_0}-1)\E\left\{\Pr\left[(\tilde{\xv}, \zv) \in T_{\epsilon}^{N}(P_{XZ})\right]\Pr\left[\tilde{\xv} \in B(w)\right]\right\}\\
   &\le \epsilon + 2^{NM(R_0-R_s)}2^{-N\left[\sum_{i=1}^{M}I(X;Z | g_i)-\epsilon\right]} \\
   &\le \epsilon + 2^{-N(\delta-\epsilon)}.
 \end{align*}
By choosing $\delta > \epsilon$, we have
\begin{equation}\label{ebound2}
    \E_{C \in \Cc}[\Pr(\mathcal{E}_2 | \pv, C] \leq \epsilon_2
\end{equation}
for every channel pair $\pv \in \Pc_{\ast}$ when the codeword length $N$ is
sufficiently large, where $\epsilon_2=\epsilon + 2^{-N(\delta-\epsilon)}$.

Now we define an error event $\mathcal{E}$, which occurs whenever
$\mathcal{E}_1$ or $\mathcal{E}_2$ occurs, i.e.
\begin{equation}\label{errorevent}
 \mathcal{E} \triangleq \mathcal{E}_1 \cup \mathcal{E}_2.
\end{equation}
According to (\ref{ebound1}) and (\ref{ebound2}), by using the union
bound, we have for any $\pv \in \Pc_{\ast}$,
\begin{eqnarray*}
\E_{C \in \Cc}[\Pr(\mathcal{E} | \pv, C)]
 &\leq&  \E_{C \in \Cc}[\Pr(\mathcal{E}_1 | \pv, C)] + \E_{C \in \Cc}[\Pr(\mathcal{E}_2 | \pv, C)]\\
 &\leq& \epsilon_1 + \epsilon_2 =\epsilon_3.
\end{eqnarray*}
It is clear that the average error probability, averaged over the channel set
$\Pc_{\ast}$ is
\begin{equation*}
\E_{\pv \in \Pc_{\ast}}\left[\E_{C \in \Cc}[\Pr(\mathcal{E} | \pv, C)]\right]
\leq \epsilon_3.
\end{equation*}
Interchanging expectations with respect to $\pv \in \Pc_{\ast}$ and with
respect to $C \in \Cc$ (since the integrand is nonnegative and bounded by 1)
yields
\begin{equation*}
    \E_{C \in \Cc}\left[\E_{\pv \in \Pc_{\ast}}[\Pr(\mathcal{E} | \pv, C)]\right] \leq
    \epsilon_3.
\end{equation*}
Then, there exists a sequence of codes $C^{\ast} \in \Cc$ (for
increasing $N$) such that
\begin{equation*}
    \E_{\pv \in \Pc_{\ast}}[\Pr(\mathcal{E} | \pv, C^{\ast})] \leq \epsilon_3,
\end{equation*}
where $\Pr(\mathcal{E} | \pv, C^{\ast})$ is a random variable that is a
function of the channel pair $\pv$. According to the Markov inequality, we have
\begin{equation*}
\Pr\left(\Pr(\mathcal{E} | \pv, C^{\ast}) \geq \sqrt{\epsilon_3} \right) \leq
\frac{\E_{\pv \in \Pc_{\ast}}[\Pr(\mathcal{E} | \pv,
C^{\ast})]}{\sqrt{\epsilon_3}} \leq \frac{\epsilon_3}{\sqrt{\epsilon_3}} =
\sqrt{\epsilon_3}.
\end{equation*}
By letting $\sqrt{\epsilon_3} = \epsilon_4$ ($\epsilon_4$ is still
arbitrarily small), we obtain that, for any $\pv \in \Pc_{\ast}$,
\begin{align}
&  &\Pr\left(\Pr(\mathcal{E} | \pv, C^{\ast}) \geq \epsilon_4\right) & \leq
\epsilon_4 & \notag\\
&\text{or} & \Pr\left(\Pr(\mathcal{E} | \pv, C^{\ast}) < \epsilon_4\right)
&\geq 1-\epsilon_4. &
\end{align}
Since $\Pr(\mathcal{E}_1 | \pv, C^{\ast})$ and $\Pr(\mathcal{E}_2 | \pv,
C^{\ast})$ are both upper bounded by $\Pr(\mathcal{E} | \pv, C^{\ast})$, we
have that
\begin{align}
&    &    \Pr\left(\Pr(\mathcal{E}_1 | \pv, C^{\ast}) < \epsilon_4\right) &\geq
1-\epsilon_4 & \label{eB1}\\
&\text{and} &    \Pr\left(\Pr(\mathcal{E}_2 | \pv, C^{\ast}) <
\epsilon_4\right) &\geq 1-\epsilon_4. &\label{eB2}
\end{align}

According to (\ref{eB1}), there exists a (non-random) sequence of codes
$C^{\ast} \in \Cc(R_0, R_s, MN)$, which when used, the legitimate receiver can
decode the message with arbitrarily small error probability for all $\pv \in
\Pc_{\ast}$ with probability 1. Inequality (\ref{eB2}) will be used in the
equivocation calculation as followed.

\subsubsection{Equivocation Calculation}

Now we calculate the equivocation rate to check whether the perfect secrecy
requirement can be satisfied when codebook $C^{\ast}$ is used.

We bound the equivocation at the eavesdropper as follows:
\begin{eqnarray*}
  H(W|\Zv, \hv,\gv)
  &=& H(W, \Zv |\hv,\gv) - H(\Zv |\hv,\gv)\\
  &=& H(W, \Zv, \Xv |\hv,\gv) - H(\Zv |\hv,\gv) - H(\Xv|W,\Zv,\hv,\gv)\\
  &=& H(\Xv |\hv,\gv) + H(W, \Zv | \Xv, \hv,\gv) - H(\Zv |\hv,\gv) - H(\Xv|W,\Zv,\hv,\gv)\\
  &\ge& H(\Xv |\hv,\gv) - I(\Xv; \Zv |\hv,\gv) -
  H(\Xv|W,\Zv,\hv,\gv).
\end{eqnarray*}
For the first term, we notice that
\begin{equation}\label{equv_T1}
H(\Xv |\hv,\gv) = NMR_0.
\end{equation}
To bound the second term, we define
$$
\mu(\Xv,\Zv|\hv,\gv) = \left\{ \begin{array}{rl}
 1 &\mbox{ if $(\Xv,\Zv) \notin T_{\epsilon}^{N}(P_{XZ})$} \\
  0 &\mbox{ otherwise.}
       \end{array} \right.
$$
Now
\begin{eqnarray}
 I(\Xv; \Zv |\hv,\gv) &\leq& I(\Xv, \mu; \Zv |\hv,\gv)      \nonumber \\
   &=& I(\Xv; \Zv |\hv,\gv,\mu) + I(\mu; \Zv |\hv,\gv) \nonumber \\
   &=& \sum_{j=0}^{1}\Pr(\mu=j)I(\Xv; \Zv |\hv,\gv,\mu=j) + I(\mu; \Zv |\hv,\gv).
\end{eqnarray}
Note that $I(\mu; \Zv |\hv,\gv) \leq h(\mu) \leq 1$,
\begin{eqnarray}
 \Pr(\mu=1)I(\Xv; \Zv |\hv,\gv,\mu=1) &\leq& N\Pr\left[(\Xv,\Zv) \notin T_{\epsilon}^{N}(P_{XZ})|\hv,\gv\right]\log_2|Z|      \nonumber \\
   &\leq& N\epsilon\log_2|Z|,\nonumber
\end{eqnarray}
and
\begin{eqnarray}
 \Pr(\mu=0)I(\Xv; \Zv |\hv,\gv,\mu=0) &\leq& I(\Xv; \Zv |\hv,\gv,\mu=0)  \nonumber \\
   &=& H(\Xv |\hv,\gv,\mu=0) + H(\Zv |\hv,\gv,\mu=0) -H(\Xv, \Zv |\hv,\gv,\mu=0)  \nonumber\\
   &\leq& N\left[MH(X) + \sum_{i=1}^{M}H(Z|g_i) - \sum_{i=1}^{M} H(X,Z |g_i) + 3\epsilon \right]  \nonumber\\
   &=& N\left[\sum_{i=1}^{M}I(X;Z | g_i) + 3\epsilon\right]
   \nonumber.
\end{eqnarray}
Therefore, we can bound the second term as
\begin{eqnarray}\label{equv_T2}
 I(\Xv; \Zv |\hv,\gv) &\leq&  N\left[\sum_{i=1}^{M}I(X;Z | g_i) + (\log_2|Z|+3)\epsilon \right] + 1  \nonumber \\
   &=& NM[R_0-R_s+\delta-(\log_2|Z|+3)\epsilon-1/N] \nonumber \\
   &=& NM(R_0-R_s+\delta_1).
\end{eqnarray}

To bound the third term, we need to use (\ref{eB2}), according to
which the eavesdropper can decode $\Xv$ with arbitrarily small error
probability, given that $W$ is known in prior and $\Zv$ is observed.
Fano's inequality implies that
\begin{equation}\label{equv_T3}
    H(\Xv|W,\Zv,\hv,\gv) \leq 1 + NM(R_0-R_s)\Pr(\mathcal{E}_2|\pv,C^{\star}) \triangleq
    NM\delta_2
\end{equation}
for every channel pair $\pv \in \Pc_{\ast}$.

Now we can combine (\ref{equv_T1}), (\ref{equv_T2}) and (\ref{equv_T3}) into
the equivocation calculation:
\begin{eqnarray}\label{equi_T}
    H(W|\Zv, \hv, \gv) &\geq& NMR_0 - NM(R_0-R_s+\delta_1) - NM\delta_2 \nonumber \\
      &=& NM(R_s-\delta_3).
\end{eqnarray}

Note that the above equivocation calculation is obtained when (non-random) code
$C^{\ast}$ is used, instead of the random code ensemble $\Cc(R_0,R_s,MN)$.
Equation (\ref{equi_T}) implies that the perfect secrecy requirement is met.
This, together with the error probability analysis, implies that code $C^{*}$
is good for all channel pairs $\pv \in \Pc_{\ast}$ with probability 1.
\end{IEEEproof}

\subsection{Proof of Theorem~\ref{thm:Mp}}

\begin{IEEEproof}
Now we show that code $C^{\ast}$ is also good for any channel pair
$\pv \in \Pc$. Note that for every $\pv=(\hv,\gv) \in \Pc$, there
always exists \emph{at least} a channel pair $\pv_{\ast}=
(\hv_{\ast},\gv_{\ast}) \in \Pc_{\ast}$, such that $\hv \succeq
\hv_{\ast}$ and $\gv \preceq \gv_{\ast}$. With the input $\Xv$, we
denote the outputs from the channel $(\hv,\gv)$ at the legitimate
receiver and the eavesdropper by $\Yv$ and $\Zv$, respectively. We
also denote by $\Yv_1$ and $\Zv_1$ the outputs at the
corresponding receivers from $(\hv_{\ast},\gv_{\ast})$. Since code
$C^{\ast}$ is good for $(\hv_{\ast},\gv_{\ast})$, $\Yv_1$ can be
decoded with arbitrarily small error probability at the legitimate
receiver and the equivocation at the eavesdropper with $\Zv_1$
being observed satisfies
\begin{equation}\label{equivvirtual}
    H(W|\Zv_{1},\gv_{*}) \geq H(W) - N\epsilon
\end{equation}
for all $\epsilon>0$ and sufficiently large $N$. Since $\hv \succeq
\hv_{\ast}$, $\Yv_1$ is a degraded version of $\Yv$, and thus if $\Yv_1$ can be
decoded at the legitimate receiver with arbitrarily small error probability,
then so can $\Yv$. We also have that
\begin{eqnarray*}
 \nonumber \lefteqn{H(W|\Zv,\gv)-H(W|\Zv_1,\gv_{\ast}) }\\
  \nonumber   &=& I(W;\Zv_{1}|\gv_{\ast})-I(W;\Zv |\gv) \geq 0,
\end{eqnarray*}
where we use the fact that $\Zv$ is a degraded version of $\Zv_1$,
since $\gv \preceq \gv_{\ast}$. Therefore,
\begin{equation}
    H(W|\Zv,\gv) \geq H(W|\Zv_{1},\gv_{\ast}) \geq H(W) -
    N\epsilon,
\end{equation}
for all $\epsilon>0$ and sufficiently large $N$, which is the perfect secrecy
requirement.
\end{IEEEproof}

\section{Proof of Theorem~\ref{thm:INR}}\label{app:inr}

\begin{IEEEproof}
We note that the punctured code $C_m$ is obtained by taking the
first $m$ blocks, $\xv(m)=[x_1^{N},\dots,x_m^{N}]$, of the mother
code $C$, where the block $x_i^N$ is transmitted over a wire-tap
channel with channel pairs $(h_i,g_i)$, for $i=1,\dots,m$. Based
on the equivalent $M$-parallel channel model, we can form a new
sequence of channel pairs by adding other $M-m$ dummy memoryless
channels whose outputs are independent of the input. For example,
we can let $h_i=0$ and $g_i=0$ for all $i = m+1, \dots, M$. The
dummy channel pairs have zero mutual information between the input
and output; that is,
\begin{align*}
& &  \sum_{i=1}^{M}I(X;Y | h_i)  &=  \sum_{i=1}^{m}I(X;Y | h_i) &\\
&\text{and} & \sum_{i=1}^{M}I(X;Z | g_i) & =  \sum_{i=1}^{m}I(X;Z | g_i).&
\end{align*}
Now, by using Theorem~\ref{thm:Mp} and the fact
$\Pc(m)\subseteq\Pc$, we have the desired result.
\end{IEEEproof}

\section{Proof of Lemma~\ref{lem:large}} \label{app:limitlm2}

Applying the weak law of large numbers, we have the following lemma that is
used in the proofs of Lemma~\ref{lem:large} and Theorem~\ref{thm:lim}.
\begin{lemmaC1}
Let $A_i$ be i.i.d. random variables with means $\mu_A$ and variances
$\sigma_A^2$. Then, for alll $\epsilon> 0$,
\begin{align}
&  &  \lim_{M\rightarrow \infty} \Pr \left[\frac{1}{M} \sum_{i=1}^{M}(A_i-\mu_A) < \epsilon \right] &= 1&  \nonumber \\
&\text{and}&  \lim_{M\rightarrow \infty} \Pr \left[\frac{1}{M}
\sum_{i=1}^{M}(A_i-\mu_A) < -\epsilon \right] &= 0.& \label{lc31}
\end{align}
\end{lemmaC1}

Now, we consider the proof of Lemma~\ref{lem:large}.

\begin{IEEEproof}
Define $A_i=(1/2)\log_2(1+\lambda_i)$ and its mean
$\mu_A=\E[A_i]$, and $B_i=(1/2)\log_2(1+\nu_i)$ and its mean
$\mu_B=\E[B_i]$, for $i=1,\dots, M$. The connection outage
probability $P_e^{[\rm INR]}$, defined in (\ref{eq:def-pe}), can
be rewritten as follows:
\begin{eqnarray*}
P_e^{[\rm INR]}
&=& \Pr\left(\frac{1}{M}\sum_{i=1}^{M}A_i < R_0\right)\\
    &=&  \Pr\left(\frac{1}{M}\sum_{i=1}^{M}(A_i-\mu_A) < R_0-\mu_A\right).
\end{eqnarray*}
By using Lemma~C.1, we have, for all $\epsilon>0$,
\begin{align}
\lim_{M\rightarrow\infty} P_e^{[\rm INR]}=\left\{
                                \begin{array}{lc}
                                  0, &  R_0 \le \mu_{A}-\epsilon\\
                                  1, &  R_0 \ge \mu_{A}+\epsilon.
                                \end{array}
                              \right.\label{eq:R00}
\end{align}

We first prove the sufficiency given by (\ref{eq:sec-re}) in
Lemma~\ref{lem:large} and show that if
\begin{equation} \label{eq:R00S}
R_0 \leq \mu_{A}-\epsilon \quad \text{and} \quad   R_0-R_s \ge R_0
\left(\frac{\mu_B}{\mu_A-\epsilon} +\epsilon\right),
\end{equation}
then (\ref{eq:sec-re}) holds.

Define
\begin{align}
M_1&=\left\lfloor \frac{MR_0}{\mu_A-\epsilon}
\right\rfloor. \label{eq:defM1}
\end{align}
Note that (\ref{eq:R00S}) implies that $M_1 \leq M$. Hence, we can bound the
secrecy outage probability $P_s^{[\rm INR]}$,  defined in (\ref{eq:def-ps}), as
follows:
\begin{align}
P_s^{[\rm INR]}&= \sum_{m=1}^{M_1} p[m]
\Pr\left(\frac{1}{M}\sum_{i=1}^{m}B_i \geq R_0-R_s\right) +
\sum_{m=M_1+1}^{M} p[m] \Pr\left(\frac{1}{M}\sum_{i=1}^{m}B_i \geq R_0-R_s\right) \nonumber\\
   &\le  \left(\sum_{m=1}^{M_1} p[m]\right) \Pr\left(\frac{1}{M}\sum_{i=1}^{M_1}B_i \geq R_0-R_s\right) + \sum_{m=M_1+1}^{M}p[m]\nonumber \\
   &\le  \Pr\left[\sum_{i=1}^{M_1}B_i \geq M(R_0-R_s)\right] + \Pr\left(\sum_{i=1}^{M_1}A_i < MR_0\right) \nonumber \\
   &=  \Pr\left[\sum_{i=1}^{M_1}\frac{B_i-\mu_B}{M_1} \geq \frac{M(R_0-R_s)}{M_1}-\mu_B\right] +
   \Pr\left(\sum_{i=1}^{M_1}\frac{A_i-\mu_A}{M_1} < \frac{MR_0}{M_1}-\mu_A\right)\notag\\
   &\le  \Pr\left[\sum_{i=1}^{M_1}\frac{B_i-\mu_B}{M_1} \ge \epsilon(\mu_A-\epsilon)\right] +
   \Pr\left(\sum_{i=1}^{M_1}\frac{A_i-\mu_A}{M_1} < \frac{MR_0}{M_1}-\mu_A\right) \label{eq:sop_ub}
  \end{align}
where the last step follows from the condition (\ref{eq:R00S}) and
the definition of $M_1$ in (\ref{eq:defM1}). Applying Lemma~C.1, we
have
\begin{align}
\lim_{M\rightarrow\infty}\Pr\left[\sum_{i=1}^{M_1}\frac{B_i-\mu_B}{M_1}
\ge \epsilon(\mu_A-\epsilon)\right]=0 \label{eq:sop_ub-1}
\end{align}
and
\begin{align}
\lim_{M\rightarrow\infty}
\left(\sum_{i=1}^{M_1}\frac{A_i-\mu_A}{M_1}
<\frac{MR_0}{M_1}-\mu_A\right)
&=\lim_{M\rightarrow\infty}\Pr\left(\sum_{i=1}^{M_1}\frac{A_i-\mu_A}{M_1}
< -\epsilon\right)\notag\\
&=0.  \label{eq:sop_ub-2}
\end{align}
Combining (\ref{eq:R00}), (\ref{eq:sop_ub}), (\ref{eq:sop_ub-1}),
and (\ref{eq:sop_ub-2}), we have (\ref{eq:sec-re}).

Next, we prove the necessity given by (\ref{eq:sec-re1}) in
Lemma~\ref{lem:large}. Based on (\ref{eq:R00}) we need only to show
that if
\begin{equation}\label{eq:R00S2}
    R_0 - R_s \le  R_0 \left(\frac{\mu_B}{\mu_A} - \epsilon\right)
\quad    \text{and} \quad  R_0 < \mu_A+\epsilon,
\end{equation}
then $\lim_{M\rightarrow \infty}P_s^{[\rm INR]} = 1$. Define
\begin{align}
M_2=\left\lceil \frac{M(R_0-R_s)}{\mu_B-\epsilon_2}\right\rceil
\end{align}
where $\epsilon_2=(\mu_A-\epsilon)\epsilon$. Note that the condition
(\ref{eq:R00S2}) implies that $M_2 \leq M$. In this case, we obtain
the following lower bound on $P_s^{[\rm INR]}$:
\begin{align}
P_s^{[\rm INR]} 
   &\ge  \sum_{m=M_2}^{M} p[m] \Pr\left[\sum_{i=1}^{m}B_i \geq M(R_0-R_s)\right] \nonumber\\
   &\ge  \left(\sum_{m=M_2}^{M} p[m]\right) \Pr\left[\sum_{i=1}^{M_2}B_i \geq M(R_0-R_s)\right] \nonumber \\
   &=  \Pr\left(\sum_{i=1}^{M_2-1}A_i < MR_0\right) \Pr\left[\sum_{i=1}^{M_2}B_i
   \ge M(R_0-R_s)\right] \nonumber \\
   &=  \Pr\left(\sum_{i=1}^{M_2-1}\frac{A_i-\mu_A}{M_2-1} < \frac{MR_0}{M_2-1}-\mu_A\right)
   \Pr\left[\sum_{i=1}^{M_2}\frac{B_i-\mu_B}{M_2} \geq
   \frac{M(R_0-R_s)}{M_2}-\mu_B\right].
   \label{eq:sop_lb}
\end{align}
Based on the condition (\ref{eq:R00S2}) and the definitions of $M_2$
and $\epsilon_2$, we have
\begin{eqnarray*}
   \frac{MR_0}{M_2-1}-\mu_A &=&  \frac{MR_0}{\lceil M(R_0-R_s)/(\mu_B-\epsilon_2) \rceil -1
   }-\mu_A \\
   &\geq& \frac{R_0}{R_0-R_s}(\mu_B - \epsilon_2)- \mu_A \\
   &\ge&\frac{\mu_A\epsilon^2}{\mu_B-\epsilon \mu_A}\\
   &>&0.
  \end{eqnarray*}
By applying Lemma~C.1, we have
\begin{align}
\lim_{M\rightarrow\infty}\Pr\left(\sum_{i=1}^{M_2-1}\frac{A_i-\mu_A}{M_2-1}
< \frac{MR_0}{M_2-1}-\mu_A\right)=1.\label{eq:sop_lb-1}
\end{align}
On the other hand, since
\begin{align*}
\frac{M(R_0-R_s)}{M_2}-\mu_B \leq -\epsilon_2 <0,
\end{align*}
Lemma~C.1 implies that
\begin{equation}
\lim_{M\rightarrow\infty}
\Pr\left(\frac{1}{M_2}\sum_{i=1}^{M_2}(B_i-\mu_B) \geq
    \frac{R_0-R_s}{M_2}-\mu_B\right) = 1. \label{eq:sop_lb-2}
\end{equation}
Finally, combining (\ref{eq:R00}), (\ref{eq:sop_lb}),
(\ref{eq:sop_lb-1}), and (\ref{eq:sop_lb-2}), we have the necessity
of Lemma~\ref{lem:large}.
\end{IEEEproof}

\section{Proof of Theorem~\ref{thm:lim}}\label{app:RTD}

To derive Theorem~\ref{thm:lim}, we need the following lemmas from
\cite{Caire:IT:01}.

\begin{lemmaD1}\label{lem:d1}
Suppose $A$ be a random variable with CDF $F_A$. Then, for all $a$ and
$\tilde{a}$. we have
\begin{equation}
    F_A(a) \leq F_A(\tilde{a})+{\bf 1}(a\geq \tilde{a})
\end{equation}
where ${\bf 1}(\cdot)$ denote the indicator function.
\end{lemmaD1}

\begin{lemmaD2}\label{lem:d2}
Suppose \{$A_i\}$ is a sequence of i.i.d. zero mean random variables with
variances $\sigma_A^2$. Then, for all $\epsilon > 0$ and sufficiently large
$n$,
\begin{equation}
   \Pr\left(\frac{1}{\sqrt{n}} \sum_{i=1}^{n}A_i < -\sqrt{n} \epsilon
   \right) \leq \exp\left(-n\frac{\epsilon^2}{2\sigma_{A}^{2}}\right).
\end{equation}
\end{lemmaD2}
We note that Lemma~D.2 follows from the central limit theorem and the bound on
the Gaussian tail function, $Q(a) \leq \exp(-a^2/2)$, where $Q$ denotes the
tail function of the standard Gaussian distribution.

\subsection{INR Protocol}
\begin{IEEEproof}
Again, we define $A_i=(1/2)\log_2(1+\lambda_i)$ with mean
$\mu_A=\E[A_i]$ and variance $\sigma_{A}^2$,
$B_i=(1/2)\log_2(1+\nu_i)$ with mean $\mu_B=\E[B_i]$, for
$i=1,\dots, M$, and
$$M_4=\left\lfloor \frac{MR_0}{\mu_A+\epsilon} \right\rfloor.$$
The reliability condition in (\ref{eq:R00}) implies $M_4\le M$.

We first consider an upper bound of $\eta^{[\rm INR]}$ based on
(\ref{eta_asmp0}):
\begin{eqnarray}
\eta^{[\rm INR]}
   & \le & MR_s \left[\sum_{m=1}^{M_4}\Pr\left(\sum_{i=1}^{m}A_i < MR_0\right)\right]^{-1} \nonumber\\
   & \le & MR_s \left[\sum_{m=1}^{M_4}\Pr\left(\sum_{i=1}^{M_4}A_i < MR_0\right)\right]^{-1} \nonumber\\
   & = & \frac{MR_s}{M_4} \left\{\Pr\left[\sum_{i=1}^{M_4}\frac{A_i-\mu_A}{M_4} <
   \frac{MR_0}{M_4} -\mu_A\right]\right\}^{-1} \nonumber.
  \end{eqnarray}
Since $MR_0/M_4-\mu_A \geq \epsilon >0$, according to Lemma~C.1, we
have
\begin{align}
\lim_{M\rightarrow\infty}\Pr\left[\sum_{i=1}^{M_4}\frac{A_i-\mu_A}{M_4}
< \frac{MR_0}{M_4} -\mu_A\right]=1.
\end{align}
Hence,
\begin{equation}\label{oneovereta1b}
\lim_{M\rightarrow\infty} \eta^{[\rm INR]} \le
\lim_{M\rightarrow\infty}\frac{MR_s}{M_4} = \frac{R_s}{R_0}\mu_A.
\end{equation}

Next, we consider a lower bound on $\eta^{[\rm INR]}$. Let
$M_5=\lfloor MR_0/(\mu_A - \epsilon) \rfloor$. We have
\begin{eqnarray}
\frac{1}{\eta^{[\rm INR]}}
   & \leq & \frac{1}{MR_s}+\frac{1}{MR_s} \sum_{m=1}^{M}
   \left[\Pr\left(\sum_{i=1}^{m}\frac{A_i}{m} <\mu_A - \epsilon\right) + {\bf 1}{\left(\frac{MR_0}{m} \ge \mu_A - \epsilon\right)}\right] \label{uselmC1}\\
   & = & \frac{1+L(M)}{MR_s} + \frac{M_5}{MR_s},
   \nonumber
  \end{eqnarray}
where (\ref{uselmC1}) follows from Lemma~D.1 and
\begin{align}
L(M)=\sum_{m=1}^{M} \Pr\left(\frac{1}{m}\sum_{i=1}^{m}(A_i-\mu_A) <
- \epsilon \right).
\end{align}
By Lemma~D.2, there exists an integer $n$, finite and independent of
$R_0$, so that
\begin{eqnarray*}
L(M)&=& \sum_{m=1}^{n} \Pr\left(\frac{1}{m}\sum_{i=1}^{m}(A_i-\mu_A)
< - \epsilon \right) + \sum_{m=n+1}^{M}
\Pr\left(\frac{1}{m}\sum_{i=1}^{m}(A_i-\mu_A) < -
\epsilon \right)\\
   &\leq& \sum_{m=1}^{n}
\Pr\left(\frac{1}{m}\sum_{i=1}^{m}(A_i-\mu_A) < - \epsilon \right)
  + \sum_{m=n+1}^{\infty}
  \exp\left(-m\frac{\epsilon^2}{2\sigma_{A}^{2}}\right).
\end{eqnarray*}
Since the first sum contains a finite number of terms (each being less than
$1$), and the second converges for all $\epsilon>0$, we have that
\begin{equation*}
    \lim_{M\rightarrow\infty} \frac{1+L(M)}{MR_s} =0.
\end{equation*}
Hence, we have that
\begin{equation}\label{oneovereta2b}
\lim_{M\rightarrow \infty} \eta^{[\rm INR]} \ge \lim_{M\rightarrow
\infty}\frac{MR_s}{M_5} = \frac{R_s}{R_0}\mu_A.
\end{equation}
Combining (\ref{oneovereta1b}) and (\ref{oneovereta2b}), we obtain
\begin{equation}
\lim_{M\rightarrow \infty} \eta^{[\rm INR]} = \frac{R_s}{R_0}\mu_A=
  \frac{R_s}{2R_0}\E\left[\log_2(1+\lambda)\right]. \label{eq:etaM}
\end{equation}
Furthermore, Lemma~\ref{lem:large} implies that
\begin{equation}\label{eq:R0Rs}
    \frac{R_s}{R_0} \leq 1 -
    \frac{\E[\log_2(1+\nu)]}{\E[\log_2(1+\lambda)]}.
\end{equation}
Finally, combining (\ref{eq:etaM}) and (\ref{eq:R0Rs}), we have the
desired result that
\begin{equation*}
    \lim_{M\rightarrow \infty} \eta^{[\rm INR]} =
    \frac{1}{2}\E\left[ \log_2(1+\lambda)- \log_2(1+\nu)\right].
\end{equation*}
\end{IEEEproof}

\subsection{RTD Scheme}
\begin{IEEEproof}
We first consider the connection outage probability $P_e^{[\rm
RTD]}$. Let $A_i=\lambda_i$ with mean $\mu_A=\E[\lambda_i]$, for
$i=1,\dots,M$. Based on (\ref{eq:def-pe}) we have
\begin{eqnarray*}
P_e^{[\rm RTD]}
&=& \Pr\left[\frac{1}{2M}\log_2\left(1+\sum_{i=1}^{M}A_i\right)< R_0\right]\\
    &=&  \Pr\left(\sum_{i=1}^{M}\frac{A_i-\mu_A}{M} < \frac{2^{2MR_0}-1}{M}-\mu_A\right).
\end{eqnarray*}
By using Lemma~C.1, we have, for all $\epsilon>0$,
\begin{align}
\lim_{M\rightarrow\infty} P_e^{[\rm RTD]}=\left\{
                                \begin{array}{lc}
                                  0, &  \frac{1}{M}(2^{2MR_0}-1) \le \mu_{A}-\epsilon\\
                                  1, &  \frac{1}{M}(2^{2MR_0}-1)\ge \mu_{A}+\epsilon.
                                \end{array}
                              \right.\label{eq:RTDPe}
\end{align}
Hence, to ensure the connection outage requirement, $R_0$ should
satisfy
\begin{align}
\frac{2^{2MR_0}-1}{M}< \mu_{A}+\epsilon. \label{eq:lim-RTD-con}
\end{align}

Now, we consider an upper bound on $\eta^{[\rm RTD]}$. Let
$$M_3=\left \lfloor \frac{2^{2MR_0}-1}{\mu_A+\epsilon} \right\rfloor < M,$$
where the inequality follows from (\ref{eq:lim-RTD-con}). By using
(\ref{eta_asmp0}), we have
\begin{eqnarray}
\eta^{[\rm RTD]}& \le & MR_s \left[1+\sum_{m=1}^{M_3} \Pr\left(\sum_{i=1}^{m}A_i < 2^{2MR_0}-1\right)\right]^{-1} \notag\\
& \le & MR_s \left[\sum_{m=1}^{M_3} \Pr\left(\sum_{i=1}^{M_3}A_i <
2^{2MR_0}-1\right)\right]^{-1} \notag\\
& \le & \frac{MR_0}{M_3}\left[\Pr\left(\sum_{i=1}^{M_3} \frac{A_i-\mu_A}{M_3} <
\frac{2^{2MR_0}-1}{M_3}-\mu_A\right)\right]^{-1}. \nonumber
\end{eqnarray}
Since ${(2^{2MR_0}-1)}/{M_3}-\mu_A \geq \epsilon >0$ and Lemma~C.1,
we have that
\begin{equation*}
   \lim_{M\rightarrow \infty}
   \Pr\left(\sum_{i=1}^{M_3}\frac{A_i-\mu_A}{M_3} < \frac{2^{2MR_0}-1}{M_3}-\mu_A\right)
   =1.
\end{equation*}
Therefore,
\begin{equation*}
  \lim_{M\rightarrow \infty} \eta^{[\mathrm{RTD}]} \le \lim_{M\rightarrow \infty}\frac{MR_0}{M_3}
  = \lim_{M\rightarrow
  \infty}\frac{MR_0(\mu_A+\epsilon_3)}{2^{2MR_0}-1}=0.
\end{equation*}
\end{IEEEproof}

\bibliographystyle{IEEEtran}
\bibliography{secthru,arq}

\end{document}